\documentclass[aps,prd,onecolumn,groupedaddress,showpacs,nofootinbib,amssymb,10pt]{revtex4-1}
\usepackage{graphicx,bm}
\usepackage{amsmath,amssymb,amsfonts}
\usepackage{subfigure}
\usepackage{microtype}

\begin{document}

\newcommand{\cL}{{\cal L}}
\newcommand{\be}{\begin{equation}}
\newcommand{\ee}{\end{equation}}
\newcommand{\bea}{\begin{eqnarray}}
\newcommand{\eea}{\end{eqnarray}}
\newcommand{\beq}{\begin{eqnarray}}
\newcommand{\eeq}{\end{eqnarray}}
\newcommand{\tr}{{\rm tr}\, }
\newcommand{\nn}{\nonumber \\}
\newcommand{\e}{{\rm e}}
\def\rf#1{(\ref{#1})}

\newcommand{\mc}[1]{\mathcal{#1}}
\newcommand{\mr}[1]{\mathrm{#1}}
\newcommand{\rd}{\mathrm{d}}
\newcommand{\vect}[1]{\bm{#1}}

\newcommand{\lambdaw}{\Lambda_W^{}}
\newcommand{\CG}{\mathcal{G}}
\newcommand{\trace}{\mathrm{Tr}}
\newcommand{\p}{\partial}

\newcommand{\g}[1][]{g^{(3)#1}}
\newcommand{\dg}[1][]{\dot{g}^{(3)#1}}
\newcommand{\bg}[1][]{\bar{g}^{(3)#1}}
\newcommand{\dbg}[1][]{\dot{\bar{g}}^{(3)#1}}
\newcommand{\bpi}{\bar{\pi}}
\newcommand{\bcLR}{\bar{\mathcal{L}}^{(3)}_R}
\newcommand{\R}[1][]{R^{(3)#1}}

\title{Modified first-order Ho\v{r}ava-Lifshitz gravity:
Hamiltonian analysis of the general theory and accelerating FRW cosmology in power-law $F(R)$ model}

\author{
Sante Carloni$^1$,
Masud Chaichian$^{2,3}$,
Shin'ichi Nojiri$^4$, \\
Sergei D. Odintsov$^{1,5}$\footnote{Also at Tomsk State Pedagogical
University}, Markku Oksanen$^2$, and Anca Tureanu$^{2,3}$}
\affiliation{ $^1$ Institut de Ciencies de l'Espai (IEEC-CSIC),
Campus UAB, Facultat de Ciencies, Torre C5-Par-2a pl, E-08193
Bellaterra
(Barcelona), Spain \\
$^2$ Department of Physics, University of Helsinki, P.O. Box 64, FI-00014 Helsinki, Finland \\
$^3$ Helsinki Institute of Physics, P.O. Box 64, FI-00014 Helsinki, Finland \\
$^4$ Department of Physics, Nagoya University, Nagoya 464-8602, Japan \\
$^5$ Instituci\`{o} Catalana de Recerca i Estudis Avan\c{c}ats (ICREA),
Barcelona }

\begin{abstract}
We propose the most general modified first-order Ho\v{r}ava-Lifshitz gravity,
whose action does not contain time derivatives higher than the second order.
The Hamiltonian structure of this theory is studied in all the details in the case
of the spatially-flat FRW space-time, demonstrating many of the features of
the general theory. It is shown that, with some plausible assumptions, 
including the projectability of the lapse function, this model is consistent. 
As a large class of such theories, the modified Ho\v{r}ava-Lifshitz 
$F(R)$ gravity is introduced. The study of its ultraviolet properties
shows that its $z=3$ version seems to be renormalizable in the same way
as the original Ho\v{r}ava-Lifshitz proposal. The Hamiltonian analysis of the
modified Ho\v{r}ava-Lifshitz $F(R)$ gravity shows that it is in general a
consistent theory. The $F(R)$ gravity action is also studied in the fixed-gauge
form, where the appearance of a scalar field is particularly illustrative. Then
the spatially-flat FRW cosmology for this $F(R)$ gravity is investigated. It is
shown that a special choice of parameters for this theory leads to the same
equations of motion as in the case of traditional $F(R)$ gravity. Nevertheless,
the cosmological structure of the modified Ho\v{r}ava-Lifshitz $F(R)$ gravity
turns out to be much richer than for its traditional counterpart. The emergence
of multiple de Sitter solutions indicates to the possibility of unification of
early-time inflation with late-time acceleration within the same model.
Power-law $F(R)$ theories are also investigated in detail. It is analytically
shown that they have a quite rich cosmological structure: early/late-time
cosmic acceleration of quintessence, as well as of phantom types. Also it is
demonstrated that all the four known types of finite-time future singularities
may occur in the power-law Ho\v{r}ava-Lifshitz $F(R)$ gravity. Finally, a covariant
proposal for (renormalizable) $F(R)$ gravity within the Ho\v{r}ava-Lifshitz
spirit is presented.
\end{abstract}


\maketitle

\section{Introduction}

Recently, it has become clear that our universe has not only
undergone the period of early-time accelerated expansion
(inflation), but also is currently in the so-called late-time
accelerating epoch (dark energy era). An extremely powerful way to
describe the early-time inflation and the late-time acceleration in
a unified manner is modified gravity. This approach does not require
the introduction of new dark components like inflaton and dark
energy. The unified description of inflation and dark energy is
achieved by modifying the gravitational action at the very early
universe as well as at the very late times (for a review of such
models, see \cite{Nojiri:2006ri}). A number of viable modified
gravity theories has been suggested. Despite some indications to
possible connection with string/M-theory \cite{Nojiri:2003rz}, such
theories remain to be mainly phenomenological. It is a challenge to
investigate their origin from some (not yet constructed) fundamental
quantum gravity theory.

Among the recent attempts to construct a consistent theory of
quantum gravity much attention has been paid to the quite remarkable
Ho\v{r}ava-Lifshitz quantum gravity \cite{Horava:2009uw}, which
appears to be power-counting renormalizable in four dimensions. In
this theory the local Lorentz invariance is abandoned, but it is
restored as an approximate symmetry at low energies. Despite its
partial success as a candidate for fundamental theory of gravity,
there are a number of unresolved problems related to the detailed
balance and projectability conditions, consistency, its general
relativity (GR) limit, realistic cosmological applications, the
relation to other modified gravities, etc. Due to the fact that its
spatially-flat FRW cosmology \cite{cosmology} is almost the same as
in GR, it is difficult to obtain a unified description of the
early-time inflation with the late-time acceleration in the standard
Ho\v{r}ava-Lifshitz gravity. 

Recently the modified Ho\v{r}ava-Lifshitz $F(R)$ gravity has been
proposed \cite{Chaichian:2010yi}. Such a modification may be easily
related with the traditional modified gravity approach, but turns
out to be much richer in terms of the possible cosmological
solutions. For instance, the unification of inflation with dark
energy seems to be possible in such Ho\v{r}ava-Lifshitz gravity due
to the presence of multiple de Sitter solutions. Moreover, on the
one hand, there is the hope that the generalization of
Ho\v{r}ava-Lifshitz gravity may lead to new classes of
renormalizable quantum gravity. On the other hand, one may hope to
formulate the dynamical scenario for the Lorentz symmetry
violation/restoration, caused by the expansion of the universe, in
terms of such generalized theory.

In the present work (section \ref{sec:2}) we propose the most
general modified first-order Ho\v{r}ava-Lifshitz-like theory,
without higher derivative terms which are normally responsible for
the presence of ghosts. The general form of the action in the
spatially-flat FRW space-time is found, and the Hamiltonian
structure of the action is analyzed in section \ref{sec:3}.

As a specific example of such a first-order action we introduce the
modified Ho\v{r}ava-Lifshitz $F(R)$ theory which is more general
than the model of ref.~\cite{Chaichian:2010yi}. Nevertheless, its
spatially-flat FRW cosmology turns out to be the same as for the
model \cite{Chaichian:2010yi} (this is not the case for black hole
solutions, etc). Therefore it also coincides with the conventional
$F(R)$ spatially-flat cosmology for a specific choice of the
parameters. The ultraviolet structure of the new Ho\v{r}ava-Lifshitz
$F(R)$ gravity is carefully investigated. It is shown that such
models can have very nice ultraviolet behaviour at $z=2$. Moreover,
for $z=3$ a big class of renormalizable models is suggested (section
\ref{sec:2}). The Hamiltonian analysis of the modified 
Ho\v{r}ava-Lifshitz $F(R)$ gravity is presented in section
\ref{sec:4}. The fixed gauge modified Ho\v{r}ava-Lifshitz 
$F(R)$ gravity is analyzed in section \ref{sec:5}.

Section \ref{sec:6} is devoted to the investigation of
spatially-flat FRW cosmology  for power-law $F(R)$ gravity. The
general equation for the de Sitter solutions is obtained. It
acquires an extremely simple form for a special choice of
parameters, when de Sitter solutions are roots of the equation
$F=0$. The existence of multiple de Sitter solutions indicates the
principal possibility of attaining the unification of the early-time
inflation with the late-time acceleration in the modified
Ho\v{r}ava-Lifshitz $F(R)$ gravity. The reconstruction technique is
developed for the study of analytical and accelerating FRW
cosmologies in power-law models. A number of explicit analytical
solutions are presented. It is shown by explicit examples that some
of the quintessence/phantom-like cosmologies may develop the future
finite-time singularity of all the known four types, precisely in
the same way as for traditional dark energy models. The possible
curing of such singularities could be achieved in a similar way as
in the case of traditional modified gravity. Some remarks about
small corrections to the Newton law are made in section \ref{sec:7}.
A summary and outlook are given in the last section \ref{sec:8}.
In the appendix \ref{appendix} we propose a covariant $F(R)$ gravity 
that is quite similar to the corresponding Ho\v{r}ava-Lifshitz version
but remains to be a covariant theory. It seems that it could also be
made renormalizable.

\section{General action for Ho\v{r}ava-Lifshitz-like gravity and
renormalizability}
\label{sec:2}

In this section we propose the essentially most general
Ho\v{r}ava-Lifshitz-like gravity action, which does not contain derivatives
with respect to the time coordinate higher than the second order.
Its ultraviolet properties are discussed.

By using the Arnowitt-Deser-Misner (ADM) decomposition \cite{Arnowitt:1962hi} (for reviews
and mathematical background, see \cite{ADMreviewmath}), one can
write the metric of space-time in the following form: 
\be \label{HLF2} 
\rd s^2 = - N^2 \rd t^2 + g^{(3)}_{ij}\left(\rd x^i +
N^i \rd t \right)\left(\rd x^j + N^j \rd t \right), \quad
i,j=1,2,3\, . \ee Here $N$ is called the lapse variable and $N^i$ is
the shift 3-vector. Then the scalar curvature $R$ has the following
form: 
\be \label{HLF3} 
R= K^{ij} K_{ij} - K^2 + R^{(3)} + 2
\nabla_\mu \left( n^\mu \nabla_\nu n^\nu
 - n^\nu \nabla_\nu n^\mu \right) \, .
\ee
Here $R^{(3)}$ is the three-dimensional scalar
curvature defined by the metric $g^{(3)}_{ij}$ and $K_{ij}$ is the
extrinsic curvature defined by
\be
\label{HLF4}
K_{ij}=\frac{1}{2N}\left(\dot
g^{(3)}_{ij}-\nabla^{(3)}_iN_j-\nabla^{(3)}_jN_i\right) \, ,\quad K
=K^i_{\ i}\, .
\ee
$n^\mu$ is the unit vector perpendicular to the
three-dimensional space-like hypersurface $\Sigma_t$ defined by $t=\text{constant}$ and
$\nabla^{(3)}_i$ is the covariant derivative on the
hypersurface $\Sigma_t$. From the determinant of the metric
\eqref{HLF2} one obtains $\sqrt{-g} = \sqrt{g^{(3)}} N$.

For general Ho\v{r}ava-Lifshitz-like gravity models, we do not
require the full diffeomorphism-invariance,  but only invariance
under ``foliation-preserving'' diffeomorphisms: 
\be \label{fpd1}
\delta x^i=\zeta^i(t,\bm{x})\,, \, \quad \delta t=f(t)\, . 
\ee
Therefore, there are many invariants or covariant quantities made
from the metric, in particular $K$, $K_{ij}$, $\nabla^{(3)}_i
K_{jk}$, $\cdots$, $\nabla^{(3)}_{i_1} \nabla^{(3)}_{i_2} \cdots
\nabla^{(3)}_{i_n} K_{jk}$, $\cdots$, $R^{(3)}$, $R^{(3)}_{ij}$,
$R^{(3)}_{ijkl}$, $\nabla^{(3)}_i R^{(3)}_{jklm}$, $\cdots$,
$\nabla^{(3)}_{i_1} \nabla^{(3)}_{i_2} \cdots \nabla^{(3)}_{i_n}
R^{(3)}_{jklm}$, $\cdots$, $\nabla_\mu \left( n^\mu \nabla_\nu n^\nu
- n^\nu \nabla_\nu n^\mu \right)$, $\cdots$, etc. Then the general
consistent action composed of invariants that are constructed from
such covariant quantities,
\bea \label{HLF26}
S_\mathrm{gHL} &=& \int \rd^4 x \sqrt{g^{(3)}} N F
\left(g^{(3)}_{ij}, K, K_{ij}, \nabla^{(3)}_i K_{jk}, \cdots,
\nabla^{(3)}_{i_1} \nabla^{(3)}_{i_2} \cdots
\nabla^{(3)}_{i_n} K_{jk}, \cdots, \right. \nn 
&& \left. R^{(3)}, R^{(3)}_{ij}, R^{(3)}_{ijkl}, \nabla^{(3)}_i
R^{(3)}_{jklm}, \cdots, \nabla^{(3)}_{i_1} \nabla^{(3)}_{i_2} \cdots
\nabla^{(3)}_{i_n} R^{(3)}_{jklm}, \cdots, \nabla_\mu \left( n^\mu
\nabla_\nu n^\nu - n^\nu \nabla_\nu n^\mu \right) \right)\, ,
\eea
could be a rather general action for the generalized
Ho\v{r}ava-Lifshitz gravity. Note that one can also include the
(cosmological) constant in the above action. Here it has been
assumed that the action does not contain derivatives higher than the
second order with respect to the time coordinate $t$. In the usual
$F(R)$ gravity, there appears the extra scalar mode, since the
equations given by the variation over the metric tensor contain the
fourth derivative. By assuming that the action does not contain
derivatives higher than the second order with respect to the time
coordinate $t$, we can avoid more extra modes in addition to the
only one scalar mode which appears in the usual $F(R)$ gravity. For
example, if we consider the action containing the terms like
\be \label{example}
\left(\nabla^\mu \nabla_\mu\right)^{n+1} R^{(3)}\,
,\quad \left(\nabla^\rho \nabla_\rho \right)^n \nabla_\mu \left(
n^\mu \nabla_\nu n^\nu - n^\nu \nabla_\nu n^\mu \right)\, ,
\ee
the equations given by the variation over the metric tensor contain
the fifth or higher derivatives (for a review of Hamiltonian structure
of higher derivative modified gravity, see \cite{Woodard:2007}). If
we define new fields recursively
\be \label{example2} 
\chi_R^{(m+1)} = \nabla^\mu \nabla_\mu \chi_R^{(m)}\, ,\quad 
\chi_R^{(0)} = R^{(3)}\, ,\quad
\chi_n^{(m+1)}  = \nabla^\mu \nabla_\mu \chi_n^{(m)}\, ,\quad
\chi_n^{(0)} = \nabla_\mu \left( n^\mu \nabla_\nu n^\nu - n^\nu
\nabla_\nu n^\mu \right) \, , 
\ee
the equations can be rewritten so
that only second derivatives appear. The scalar fields in
(\ref{example2}), however, often become ghost fields that generate
states of negative norm. Thus, we only consider actions of the form 
given by (\ref{HLF26}) in this paper.

In the Ho\v{r}ava-Lifshitz-type gravity, we assume that $N$
can only depend on the time coordinate $t$, which is called the
\emph{projectability condition}. The reason is that the
Ho\v{r}ava-Lifshitz gravity does not have the full diffeomorphism-invariance,
but is invariant only under the foliation-preserving
diffeomorphisms (\ref{fpd1}).
If $N$ depended on the spatial coordinates, we could not fix $N$ to be
unity ($N=1$) by using the foliation-preserving diffeomorphisms.
Moreover, there are strong reasons to suspect that the non-projectable
version of the Ho\v{r}ava-Lifshitz gravity is generally inconsistent \cite{nonprojectable}.
Therefore we prefer to assume that $N$ is projectable.

In the FRW space-time with the flat spatial part and the
non-trivial lapse $N(t)$,
\be
\label{HLF8}
\rd s^2 = - N(t)^2 \rd t^2 + a(t)^2 \sum_{i=1}^3 \left( \rd x^i
\right)^2\, ,
\ee
we find
\bea
\label{HLF27}
&& \Gamma^0_{00} =
\frac{\dot N}{N}\, , \quad \Gamma^0_{ij} = \frac{a^2
H}{N^2}\delta_{ij}\, , \quad \Gamma^i_{j0} = H\delta^i_{\ j}\, \quad
\mbox{other}\ \Gamma^\mu_{\nu\rho} = 0\, , \nn
&& K_{ij} =
\frac{a^2H}{N}\delta_{ij}\, ,\quad \nabla^{(3)}_i = 0\, , \quad
R^{(3)}_{ijkl}=0\, ,\quad \nabla_\mu \left( n^\mu \nabla_\nu n^\nu -
n^\nu \nabla_\nu n^\mu \right) = \frac{3}{a^3 N}\frac{\rd}{\rd
t}\left(\frac{a^3 H}{N}\right)\, ,
\eea
where $H=\frac{\dot{a}}{a}$ is the Hubble parameter.
Then one gets
\bea
\label{HLF27b}
&& g^{(3)}_{ij}=a^2\delta_{ij}\, , \quad K=\frac{3
H}{N}\, , \quad K_{ij}K^{ij}=3\left(\frac{H}{N}\right)^2\, ,\quad
\nabla^{(3)}_i K_{jk}= \cdots = \nabla^{(3)}_{i_1}
\nabla^{(3)}_{i_2} \cdots \nabla^{(3)}_{i_n} K_{jk} = \cdots = 0 \,
,\nn
&& R^{(3)}=R^{(3)}_{ij}=R^{(3)}_{ijkl}=\nabla^{(3)}_i R_{jklm}=
\cdots = \nabla^{(3)}_{i_1} \nabla^{(3)}_{i_2} \cdots
\nabla^{(3)}_{i_n} R^{(3)}_{jklm} = \cdots =0\, ,
\eea
and since $F$ must be a scalar under the spatial
rotation, the action (\ref{HLF26}) reduces to
\bea
\label{HLF28}
S_\mathrm{gHL} &=& \int \rd^4 x \sqrt{g^{(3)}} N F
\left(\frac{H}{N}, \frac{3}{a^3 N}\frac{\rd}{\rd t}\left(\frac{a^3
H}{N}\right)\right)\, .
\eea
Therefore, if we consider the FRW
cosmology, the function $F$ should depend on only two variables,
$\frac{H}{N}$ and $\frac{3}{a^3 N}\frac{\rd}{\rd t}\left(\frac{a^3
H}{N}\right)$.

As a specific example of the above general theory, we may consider 
the following modified Ho\v{r}ava-Lifshitz $F(R)$ gravity, 
whose action is given by 
\be \label{HLF11}
S_{F(\tilde R)} = \frac{1}{2\kappa^2}\int \rd^4 x \sqrt{g^{(3)}} N
F(\tilde R)\, , \quad \tilde R \equiv K^{ij} K_{ij} - \lambda K^2 +
2 \mu \nabla_\mu \left( n^\mu \nabla_\nu n^\nu - n^\nu \nabla_\nu
n^\mu \right) - \mathcal{L}_R^{(3)}\left(g^{(3)}_{ij}\right) \, .
\ee 
Here $\lambda$ and $\mu$ are constants and $\mathcal{L}_R^{(3)}$
is a function of the three-dimensional metric $g^{(3)}_{ij}$ and the
covariant derivatives $\nabla^{(3)}_i$ defined by this metric. Note
that this action (\ref{HLF11}) is more general than the one
introduced in ref.~\cite{Chaichian:2010yi} due to the presence of
the last term in $\tilde{R}$. We normalize $F(\tilde R)$ or redefine
$\kappa^2$ so that 
\be \label{HLF11b} 
F'(0) = 1\, . 
\ee 
In \cite{Horava:2009uw}, $\mathcal{L}_R^{(3)}$ is chosen to be 
\be
\label{HLFrg0} \mathcal{L}_R^{(3)} \left(g^{(3)}_{ij}\right) =
E^{ij}\CG_{ijkl} E^{kl}\, , 
\ee 
where $\CG_{ijk l}$ is the
``generalized De~Witt metric'' or ``super-metric'' (``metric of the
space of metric''), 
\be \label{HLF6} 
\CG^{ijkl} = \frac{1}{2}\left(
g^{(3) ik} g^{(3) jl} + g^{(3) il} g^{(3) jk} \right) - \lambda
g^{(3) ij} g^{(3) kl}\, , 
\ee 
defined on the three-dimensional
hypersurface $\Sigma_t$. $E^{ij}$ can be defined by the so called
\emph{detailed balance condition} by using an action
$W[g^{(3)}_{kl}]$ on the hypersurface $\Sigma_t$ 
\be \label{HLF7}
\sqrt{\g}E^{ij}=\frac{\delta W[g^{(3)}_{k l}]}{\delta
g^{(3)}_{ij}}\, , \ee and the inverse of $\CG^{ijkl}$ is written as
\be \label{HLF7b0} \mc{G}_{ijkl} = \frac{1}{2}\left( g^{(3)}_{ik}
g^{(3)}_{jl} + g^{(3)}_{il} g^{(3)}_{jk} \right)
 - \tilde{\lambda} g^{(3)}_{ij} g^{(3)}_{kl}\, ,\quad \tilde{\lambda} =
\frac{\lambda}{3
\lambda - 1}\, .
\ee
The action $W[g^{(3)}_{kl}]$ is assumed to be defined by the metric and
the covariant derivatives on the hypersurface $\Sigma_t$. There is an
anisotropy between space and time in the Ho\v{r}ava-Lifshitz
gravity. In the ultraviolet (high energy) region, the time
coordinate and the spatial coordinates are assumed to behave as
\be
\label{HLF7b}
\bm{x}\to b\bm{x}\, ,\quad t\to b^z t\, ,\quad
z=2,3,\cdots\, ,
\ee
under the scale transformation. In
\cite{Horava:2009uw}, $W[g^{(3)}_{kl}]$ is explicitly given for the
case $z=2$,
\be
\label{HLF7c}
W=\frac{1}{\kappa_W^2}\int
\rd^3\vect{x}\,\sqrt{\g}\left(\R-2\lambdaw\right)\, ,
\ee
and for the case $z=3$,
\be
\label{HLF7d}
W=\frac{1}{w^2}\int_{\Sigma_t}\omega_3(\Gamma)\, ,
\ee
where
\be
\label{HLF7e}
\omega_3(\Gamma) = \trace\left(\Gamma\wedge
\rd\Gamma+\frac{2}{3}\Gamma\wedge\Gamma \wedge\Gamma\right) \equiv
\varepsilon^{ijk}\left(\Gamma^{m}_{il}\p_j
\Gamma^{l}_{km}+\frac{2}{3}\Gamma^{n}_{il}\Gamma^{l}_{jm}
\Gamma^{m}_{kn}\right)\rd^3\vect{x}\, .
\ee
Here $\kappa_W$
in (\ref{HLF7c}) is a coupling constant of dimension $-1/2$ and
$w^2$ in (\ref{HLF7d}) is a dimensionless coupling constant.
A general $E^{ij}$ consist of all contributions to $W$ up to the chosen value $z$.
The original motivation for the detailed balance condition is its ability to
simplify the quantum behaviour and renormalization properties
of theories that respect it. Otherwise there is no a priori physical
reason to restrict $\mc{L}_R^{(3)}$ to be defined by (\ref{HLFrg0}).
In the following we abandon the detailed balance condition and consider
$\mc{L}_R^{(3)}$ to have a more general form, since it is not always
relevant even for the renomalizability problem.

We now investigate the renormalizability and the unitarity of the
model
(\ref{HLF11}). For this purpose, by introducing an auxiliary field
$A$,
we rewrite the action (\ref{HLF11}) in the following form:
\be
\label{HLFrg1}
S_{F(\tilde R)} = \frac{1}{2\kappa^2}\int \rd^4 x
\sqrt{g^{(3)}} N \left\{F'(A) (\tilde R - A) + F(A)\right\}\, .
\ee
For simplicity,  the following gauge condition is used:
\be
\label{HLFrg2}
N=1\, ,\quad N^i = 0\, .
\ee
Then one finds
\bea
\label{HLFrg2b}
&& \Gamma^0_{ij} = - \frac{1}{2} {\dot g}^{(3)}_{ij}\, ,\quad
\Gamma^i_{j0} = \Gamma^i_{0j} = \frac{1}{2} g^{(3) ik} {\dot
g}^{(3)}_{kj}\, ,\quad
\Gamma^{i}_{jk} = \Gamma^{(3) i}_{jk}
\equiv \frac{1}{2} g^{(3) il}\left( g^{(3)}_{lk,j} + g^{(3)}_{jl,k}
- g^{(3)}_{jk,l} \right)\, ,\nn
&& \mbox{other components of\ } \Gamma^\mu_{\nu\rho} = 0\, ,
\eea
and therefore
\be
\label{HLFrg3}
\left(n^\mu\right) = \left( 1, 0, 0, 0 \right)\, , \quad
K_{ij}=\frac{1}{2}\dot g^{(3)}_{ij}\, , \quad
\nabla_\mu \left( n^\mu \nabla_\nu n^\nu - n^\nu \nabla_\nu
n^\mu \right) = \frac{1}{2}\partial_0 \left(g^{(3) ij}{\dot
g}^{(3)}_{ij} \right)
+ \frac{1}{4}\left(g^{(3) ij}{\dot g}^{(3)}_{ij} \right)^2\, .
\ee
We define a new field by
\be
\label{varphi}
\varphi \equiv \frac{1}{3}\ln F'(A) \, ,
\ee
which can be algebraically solved as $A=A(\varphi)$, so that
\be
\label{Avarphi}
\varphi = \frac{1}{3}\ln F'(A(\varphi)) \quad\Leftrightarrow\quad
F'(A(\varphi)) = \e^{3\varphi} \, .
\ee
The spatial metric is redefined as
\be
\label{HLFrg4}
g^{(3)}_{ij} = \e^{-\varphi} {\bar g}^{(3)}_{ij}\, .
\ee
Then the action (\ref{HLFrg1}) has the following form:
\bea
\label{HLFrg1b}
S_{F(\tilde R)} &=& \frac{1}{2\kappa^2}\int \rd^4 x
\sqrt{{\bar g}^{(3)}} \left\{
\frac{1}{4}{\bar g}^{(3) ij}{\bar g}^{(3) kl}\dot{\bar
g}^{(3)}_{ik} \dot{\bar g}^{(3)}_{jl}
 - \frac{\lambda}{4} \left( {\bar g}^{(3) ij}\dot{\bar
g}^{(3)}_{ij} \right)^2 \right. \nn
&& \left. + \left( - \frac{1}{2} + \frac{3\lambda}{2} -
\frac{3\mu}{2} \right)
{\bar g}^{(3) ij}\dot{\bar g}^{(3)}_{ij} \dot\varphi
+ \left( \frac{3}{4} - \frac{9\lambda}{4} + \frac{9\mu}{2} \right)
{\dot\varphi}^2
+ \bar{\mathcal{L}}_R^{(3)}\left( {\bar g}^{(3)}_{ij}, \varphi
\right)
 - V(\varphi) \right\}\, .
\eea
Here
\be
\label{HLFrg5}
\bar{\mathcal{L}}_R^{(3)}\left({\bar g}^{(3)}_{ij}, \varphi \right)
\equiv \mathcal{L}_R^{(3)}\left( \e^{-\varphi} {\bar g}^{(3)}_{ij}
\right)\, ,\quad
V(\varphi) \equiv A\left(\varphi\right)F'
\left(A\left(\varphi\right)\right))
 - F\left(A\left(\varphi\right)\right)\, .
\ee If we insert $\varphi=1$ into the action (\ref{HLFrg1b}), the
standard Ho\v{r}ava-Lifshitz gravity emerges. On the other hand, if
we choose 
\be \label{HLFrg6} 
\mu = \lambda - \frac{1}{3} \, , 
\ee
$\dot \varphi$ decouples with ${\dot g}^{(3)}_{ij}$. When the
decoupling (\ref{HLFrg6}) is assumed and 
\be \label{HLFrg7} 
\lambda > \frac{1}{3} \, , 
\ee 
$\varphi$ becomes canonical and the theory
becomes unitary. In the case 
\be \label{HLFrg8} 
\lambda= \frac{1}{3} \, , 
\ee 
the ${\dot\varphi}^2$ term vanishes and therefore $\varphi$
becomes non-dynamical, i.e. an auxiliary field. Eq. (\ref{HLFrg6})
tells that $\mu=0$ when $\lambda=1/3$.

In order to clarify the renormalizabilty issue, we need to
explicitly construct $\mathcal{L}_R^{(3)} \left(g^{(3)}_{ij}\right)$
in (\ref{HLF11}). As a model corresponding to $z=2$ in
(\ref{HLF7b}), which is still not renormalizable, we may propose 
\be
\label{HLFrg9} 
\mathcal{L}_R^{(3)} \left(g^{(3)}_{ij}\right) = c_2
\left( R^{(3) ij}R^{(3)}_{ij} + \alpha \left(R^{(3)}\right)^2
\right)\, , 
\ee 
where $c_2$ and $\alpha$ are constants. Since the
action (\ref{HLFrg9}) induces the higher derivative terms to
contribute to the propagators, and therefore the propagators behave
as $1/\left| \bm{k} \right|^4$ in the high energy region, the
ultraviolet behavior is improved, although the theory still is not
renormalizable.

By the scale transformation (\ref{HLFrg4}), the curvatures are
transformed as
\bea
\label{HLFrg10}
R^{(3)}_{ij} &=& {\bar R}^{(3)}_{ij} + \frac{1}{2}\left( {\bar
\nabla}^{(3)}_i {\bar \nabla}^{(3)}_j \varphi
+ {\bar g}^{(3)}_{ij} {\bar \triangle}^{(3)} \varphi \right)
+ \frac{1}{4} \left( {\bar \nabla}^{(3)}_i \varphi {\bar
\nabla}^{(3)}_j \varphi
 - {\bar g}^{(3)}_{ij} {\bar g}^{(3) kl} {\bar \nabla}^{(3)}_k
\varphi {\bar \nabla}^{(3)}_l \varphi \right)\, , \nn
R^{(3)} &=& \e^\varphi \left( {\bar R}^{(3)} + 2 {\bar
\triangle}^{(3)} \varphi
 - \frac{1}{2} {\bar g}^{(3) kl} {\bar \nabla}^{(3)}_k \varphi
{\bar \nabla}^{(3)}_l \varphi \right)\, .
\eea
Here ${\bar R}^{(3)}_{ij}$, ${\bar R}^{(3)}$, ${\bar
\nabla}^{(3)}_i$, and ${\bar \triangle}^{(3)}$
are the Ricci curvature, the scalar curvature, the covariant derivative, and the
Laplacian defined by the metric ${\bar g}^{(3)}_{ij}$, respectively.
Then if we consider the perturbation from the flat background, where
$\varphi\sim 0$ due to (\ref{HLF11b}),
\be
\label{HLFrg11}
{\bar g}^{(3)}_{ij} = \delta_{ij} + {\bar h}^{(3)}_{ij}\, ,\quad
\left| {\bar h}^{(3)}_{ij} \right|,\, \left| \varphi \right| \ll 1\, ,
\ee
we find
\bea
\label{HLFrg12}
&& \int \rd^4 x \sqrt{g^{(3)}} N \bar {\mathcal{L}}_R^{(3)}\left(
{\bar g}^{(3)}_{ij}, \varphi \right) \nn
&& = \int \rd^4 x \left[ \frac{1}{4} \left\{ \partial_i \partial^k
{\bar h}^{(3)}_{kj}
+ \partial_j \partial^k {\bar h}^{(3)}_{ki} - \partial_i \partial_j
{\bar h}^{(3)k}_k
 - \triangle {\bar h}^{(3)}_{ij} \right\}
\left\{ \partial^i \partial^l {\bar h}^{(3)j}_l + \partial^j
\partial^l {\bar h}^{(3)i}_l
 - \partial^i \partial^j {\bar h}^{(3)l}_l
 - \triangle {\bar h}^{(3) ij} \right\} \right. \nn
&& \left. + \alpha \left\{ \partial_i \partial^j {\bar h}^{(3)}_{ij}
- \triangle {\bar h}^{(3)k}_k \right\}^2 + \left( - \frac{1}{2} +
4\alpha \right) \left\{ \partial_i \partial^j {\bar h}^{(3)}_{ij} -
\triangle {\bar h}^{(3)k}_k \right\} \triangle \varphi + \left(
\frac{1}{2} + 4\alpha \right) \left( \triangle \varphi \right)^2
\right]\, . \eea Therefore if one chooses 
\be \label{HLFrg13} 
\alpha = \frac{1}{8}\, , 
\ee 
$\varphi$ decouples with ${\bar h}^{(3)}_{ij}$. Eq. (\ref{HLFrg12})
shows that the propagators of
$\varphi$ and ${\bar h}^{(3)}_{ij}$ behave as $1/\left| \bm{k}
\right|^4$ in the high energy region, so that the ultraviolet
behavior is improved.

Similarly, a model corresponding to $z=3$ in (\ref{HLF7b}), which
could be power-counting renormalizable, can be obtained by choosing
\be \label{HLFrg14} \mathcal{L}_R^{(3)} \left(g^{(3)}_{ij}\right) =
c_3 \left( {\bar \nabla}^{(3) k} R^{(3) ij} {\bar \nabla}^{(3)}_k
R^{(3)}_{ij} + \frac{1}{8} {\bar \nabla}^{(3) k} R^{(3)} {\bar
\nabla}^{(3)}_k R^{(3)} \right)\, . \ee For the $z=3$ model, the
dimension of $\varphi$ vanishes and therefore all the interactions
in $\bar{\mathcal{L}}_R^{(3)}\left( {\bar g}^{(3)}_{ij}, \varphi
\right)$ and $V(\varphi)$ in (\ref{HLFrg1b}) become power-counting
renormalizable. The propagators of $\varphi$ and ${\bar
h}^{(3)}_{ij}$ behave as $1/\left| \bm{k} \right|^6$ in the high
energy region, so that the ultraviolet behavior is improved to be
renormalizable.

We have shown that by requiring (\ref{HLFrg6}) and
(\ref{HLFrg13}), the scalar field decouples with
the gravity modes in the Einstein frame. The decoupling itself is not
directly related with the renormalizability, but the decoupling
makes it much easier to discuss the renormalizability of the model.
The choice (\ref{HLFrg14}) for $\mathcal{L}_R^{(3)}$ gives the 
renormalizable model. The renormalizability does not essentially
depend on the functional form of $F(R)$.

\section{Hamiltonian analysis of the general action in the FRW space-time}
\label{sec:3}

Let us analyze the proposed general action \eqref{HLF28} of the FRW
space-time \eqref{HLF8} with the flat spatial part and the
non-trivial lapse $N=N(t)$. Introducing four auxiliary variables
$\alpha, A, \beta, B$ enables us to write the action \eqref{HLF28}
as 
\be \label{action_FRW} 
S_\mr{gHL} = \int \rd^4 x \sqrt{\g} N
\left[ \alpha \left( A - \frac{H}{N} \right) + \beta \left( B -
\frac{3}{a^3 N}\frac{\rd}{\rd t}\left(\frac{a^3 H}{N}\right) \right)
+ F \left( A, B \right) \right]  \, . 
\ee 
The variations of the action \eqref{action_FRW} with respect to $\alpha$
and $\beta$ yield
\be \label{AandB} 
A = \frac{H}{N} \quad\text{and}\quad B =
\frac{3}{a^3 N}\frac{\rd}{\rd t} \left(\frac{a^3 H}{N}\right)\, ,
\ee
respectively. Integration by parts permits the removal of the second-order time
derivative of $a$ and the time derivative of $N$, assuming the
boundary terms vanish, but with the price that $\beta$ becomes a
dynamical variable. Thus the action \eqref{action_FRW} can be
written as 
\be \label{action_FRW_final} 
S_\mr{gHL} = \int \rd^4 x
\sqrt{\g} N \left[ \alpha \left( A - \frac{H}{N} \right) + \beta B +
\frac{3\dot{\beta} H}{N^2} + F \left( A, B \right) \right] \, , 
\ee
The action \eqref{action_FRW_final} is equivalent to
\eqref{action_FRW}
and consequently to the original action \eqref{HLF28}. The advantage of
the action \eqref{action_FRW_final} over \eqref{HLF28} is the
simpler dependence on the variables $a$ and $N$, which will be
crucially important in the following Hamiltonian analysis.

For the Hamiltonian analysis of constrained systems and their
quantization we refer to the monographs 
\cite{Dirac:1964,Chaichian:1984,Gitman:1990,Henneaux:1994}.

In the Hamiltonian formalism the generalized coordinates $\g_{ij}$,
$N$, $\alpha$, $A$, $\beta$ and $B$ of the action
\eqref{action_FRW_final} have the canonically conjugated momenta
$\pi^{ij}$, $\pi_N$, $\pi_\alpha$, $\pi_A$, $\pi_\beta$ and $\pi_B$,
respectively. We consider $N$ to be projectable, $N = N(t)$, and
therefore also the momentum $\pi_N = \pi_N(t)$ is constant on the
hypersurface $\Sigma_t$ for each $t$. The Poisson brackets are
postulated in the form (equal time $t$ is understood) 
\bea
\label{PB} &&\{ \g_{ij}(\vect{x}), \pi^{kl}(\vect{y}) \} =
\frac{1}{2} \left( \delta_i^k \delta_j^l + \delta_i^l \delta_j^k
\right) \delta(\vect{x} - \vect{y}) \,,\quad \{ N, \pi_N \} = 1  \,
,\nn &&\{ \alpha(\vect{x}), \pi_\alpha(\vect{y}) \} =
\delta(\vect{x} - \vect{y}) \,, \quad \{ A(\vect{x}),
\pi_A(\vect{y}) \} = \delta(\vect{x} - \vect{y})\,,\nn &&\{
\beta(\vect{x}), \pi_\beta(\vect{y}) \} = \delta(\vect{x} -
\vect{y}) \,, \quad \{ B(\vect{x}), \pi_B(\vect{y}) \} =
\delta(\vect{x} - \vect{y})\, , 
\eea 
with all the other Poisson brackets vanishing. We are considering the
FRW metric \eqref{HLF8}
with the flat spatial part $\g_{ij} = a^2 \delta_{ij}$ and therefore
the Poisson bracket for the scale factor $a$ and the momenta
conjugate to the 3-metric takes the form 
\be \label{PB_for_a}
\int\rd^3 \vect{y}\, \{ a, \pi^{ij}(\vect{y}) \} =
\frac{\delta^{ij}}{2a} \, . 
\ee

Let us find the momenta and the primary constraints. The action
\eqref{action_FRW_final} does not depend on the time derivative of
$N$, $\alpha$, $A$ or $B$. Thus we have the primary constraints 
\be
\label{p_constraints_FRW} \Phi_1 \equiv \pi_N \approx 0\, ,\quad
\Phi_2(\vect{x}) \equiv \pi_\alpha(\vect{x}) \approx 0\, ,\quad
\Phi_3(\vect{x}) \equiv \pi_A(\vect{x}) \approx 0\, , \quad
\Phi_4(\vect{x}) \equiv \pi_B(\vect{x}) \approx 0 \, . 
\ee 
The momenta conjugated to $\beta$ and $\g_{ij}$ are
\bea
\pi_\beta &=& \frac{\delta S_\mr{gHL}}{\delta \dot{\beta}} = \frac{3a^3 H}{N} \, , \label{pi_beta}\\
\pi^{ij} &=& \frac{\delta S_\mr{gHL}}{\delta \dg_{ij}}
= \frac{a}{6} \left( -\alpha + \frac{3\dot{\beta}}{N} \right)
\delta^{ij} \, , \label{pi^ij}
\eea
respectively. The ``velocities'' $\dot\beta$ and $\dg_{ij}$ can be 
solved in terms of the canonical variables, so there are no more 
primary constraints.

Then we define the Hamiltonian 
\be \label{H} 
H = \int\rd^3 \vect{x} \left( \pi^{ij} \dg_{ij} + \pi_\beta \dot{\beta}
\right) - L =
\int\rd^3 \vect{x} N \mc{H} \, , 
\ee 
where the Lagrangian $L$ is defined by \eqref{action_FRW_final},
$S_\mr{gHL} = \int \rd t L$, 
and the so-called Hamiltonian constraint is found to be 
\be 
\mc{H} = \frac{\pi_\beta}{3} \left(
\frac{2}{a} \sum_{i=1}^3 \pi^{ii} + \alpha \right) - a^3 \left(
\alpha A + \beta B + F(A, B) \right) \, . 
\ee 
The primary constraints \eqref{p_constraints_FRW} can be included into
the Hamiltonian \eqref{H} by using the Lagrange multipliers $\lambda_k$,
$k=1,2,3,4$. We define the total Hamiltonian by 
\be \label{H_T_FRW}
H_T = H + \lambda_1 \Phi_1 + \sum_{n=2}^4 \int\rd^3 \vect{x}
\lambda_n(\vect{x}) \Phi_n(\vect{x}) \, . 
\ee 
Note that there is no space integral over the product $\lambda_1 \Phi_1
= \lambda_1 \pi_N$, since these variables depend only on the time
coordinate $t$.

The consistency of the system requires that every constraint has to
be preserved under time evolution. Since the Poisson brackets of the
primary constraints \eqref{p_constraints_FRW} are zero, 
\be 
\{ \Phi_k, \Phi_l \} = 0 \, ,\ k,l \in \{1,2,3,4\} \, , 
\ee 
the time evolution of the primary constraints is determined by the
Hamiltonian $H$ alone 
\be \label{p_time_evo} 
\dot{\Phi}_k = \{
\Phi_k, H_T \} = \{ \Phi_k, H \} \, ,\ k=1,2,3,4 \, . 
\ee 
Thus the following time derivatives of the primary constraints have to
vanish: 
\bea \label{primary_preserved} 
&& \dot{\Phi}_1 = \dot{\pi}_N
= \{ \pi_N, H \} = - \int\rd^3 \vect{x} \mc{H} \, ,\nn 
&& \dot{\Phi}_2 = \dot{\pi}_\alpha = \{ \pi_\alpha, H \} = N \left(
-\frac{\pi_\beta}{3} + a^3 A \right) \, ,\nn 
&& \dot{\Phi}_3 = \dot{\pi}_A = \{ \pi_A, H \} = N a^3 \left( \alpha +
\frac{\p F(A,B)}{\p A} \right) \, ,\nn && \dot{\Phi}_4 = \dot{\pi}_B =
\{ \pi_B,
H \} = N a^3 \left( \beta + \frac{\p F(A, B)}{\p B} \right) \, .
\eea 
Since none of these expressions \eqref{primary_preserved}
vanish due to the primary constraints \eqref{p_constraints_FRW}, we
must impose the secondary constraints: 
\bea \label{s_constraints_FRW} 
\Phi_0 &\equiv& \int\rd^3 \vect{x} \mc{H}
\approx 0 \, ,\nn \Phi_5(\vect{x}) &\equiv& - \frac{\pi_\beta}{3} +
a^3 A \approx 0 \, ,\nn \Phi_6(\vect{x}) &\equiv& \alpha + \frac{\p
F(A, B)}{\p A} \approx 0 \, ,\nn \Phi_7(\vect{x}) &\equiv& \beta +
\frac{\p F(A, B)}{\p B} \approx 0 \, . 
\eea 
Here the position argument $\vect{x}$ has been omitted in the right-hand
side of the local constraints. Note that neither $N$ or $a$ can be
constrained
to vanish, since they are the essential physical quantities in this
theory. Here the actual Hamiltonian constraint $\Phi_0$ is global
due to the projectability condition, $N=N(t)$. Note that the
Hamiltonian (\ref{H}) is simply this constraint multiplied by $N$,
i.e. 
\be \label{H_as_constraint} 
H = N \Phi_0 \, . 
\ee

Also the secondary constraints \eqref{s_constraints_FRW} have to be 
preserved under time evolution. The time evolution of the secondary 
constraints is
\be \label{s_time_evo}
\dot{\Phi}_m = \{ \Phi_m, H_T \} = N \{ \Phi_m, \Phi_0 \} 
+ \sum_{n=2}^4 \int\rd^3 \vect{y}\, \lambda_n(\vect{y})
\{ \Phi_m, \Phi_n(\vect{y}) \} \, ,\ m=0,5,6,7 \, ,
\ee
where we have used \eqref{H_as_constraint} and the fact that none of 
the constraints $\Phi_j, j=0,1,2,\ldots,7$ depend on the lapse $N$,
and that the secondary constraints (\ref{s_constraints_FRW}) do not 
depend on $\pi_N$.
For the global Hamiltonian constraint $\Phi_0$ we find the following 
Poisson brackets with the primary constraints \eqref{p_constraints_FRW}
\be
\label{Phi0_PBs}
\{ \Phi_0, \Phi_2(\vect{x}) \} = - \Phi_5(\vect{x}) \, ,\quad \{ \Phi_0,
\Phi_3(\vect{x}) \} = - a^3 \Phi_6(\vect{x}) \, ,\quad \{ \Phi_0, \Phi_4(\vect{x}) \}
= - a^3 \Phi_7(\vect{x}) \, ,
\ee
which all vanish due to the other secondary constraints. Thus, according to
\eqref{s_time_evo} and \eqref{Phi0_PBs}, the Hamiltonian constraint $\Phi_0$ is preserved under time evolution, $\dot{\Phi}_0 \approx 0$.
For the secondary constraint $\Phi_5$ we obtain the non-vanishing Poisson brackets
with the primary constraints \eqref{p_constraints_FRW} and the Hamiltonian constraint
$\Phi_0$:
\be
\{\Phi_5(\vect{x}), \Phi_3(\vect{y})\} = a^3 \delta(\vect{x}-\vect{y}) \, ,\quad
\{\Phi_5, \Phi_0(\vect{x})\} = - \frac{a^3 B}{3} + 3\pi_\beta A \, .
\ee
For the next secondary constraint $\Phi_6$ we obtain the non-vanishing Poisson brackets:
\bea
\{\Phi_6(\vect{x}), \Phi_2(\vect{y})\} &=& \delta(\vect{x}-\vect{y}) \, ,\nn
\{\Phi_6(\vect{x}), \Phi_3(\vect{y})\} &=& \frac{\p^2 F(A, B)}{\p A^2}\,
\delta(\vect{x}-\vect{y}) \, ,\nn
\{\Phi_6(\vect{x}), \Phi_4(\vect{y})\} &=& \frac{\p^2 F(A, B)}{\p A \p B}\,
\delta(\vect{x}-\vect{y}) \, .
\eea
For the last secondary constraint $\Phi_7$ we obtain the non-vanishing Poisson brackets:
\bea
\{\Phi_7(\vect{x}), \Phi_3(\vect{y})\} &=& \frac{\p^2 F(A, B)}{\p A \p B}\,
\delta(\vect{x}-\vect{y}) \, ,\nn
\{\Phi_7(\vect{x}), \Phi_4(\vect{y})\} &=& \frac{\p^2 F(A, B)}{\p B^2}\,
\delta(\vect{x}-\vect{y})\, ,\nn
\{\Phi_7(\vect{x}), \Phi_0\} &=& \frac{1}{3} \left( \frac{2}{a}\sum_{i=1}^3 \pi^{ii}
+ \alpha \right) \, .
\eea
Inserting all these Poisson brackets into \eqref{s_time_evo} gives the tertiary constraints:
\bea
\dot{\Phi}_5 &=& N \left( - \frac{a^3 B}{3} + 3\pi_\beta A \right)
+ \lambda_3 a^3 \approx 0 \, .\label{t_constraint1_FRW}\\
\dot{\Phi}_6 &=& \lambda_2 + \lambda_3 \frac{\p^2 F(A, B)}{\p A^2}
+ \lambda_4 \frac{\p^2 F(A, B)}{\p A \p B} \approx 0 \, .\label{t_constraint2_FRW}\\
\dot{\Phi}_7 &=& \frac{N}{3} \left( \frac{2}{a}\sum_{i=1}^3 \pi^{ii}
+ \alpha \right) + \lambda_3 \frac{\p^2 F(A, B)}{\p A \p B}
+ \lambda_4 \frac{\p^2 F(A, B)}{\p B^2} \approx 0 \, .\label{t_constraint3_FRW}
\eea
We assume that all the second partial derivatives of $F(A, B)$ do not
vanish.\footnote{This is the case for example in the modified
Ho\v{r}ava-Lifshitz gravity model $F(\tilde{R}) \propto \tilde{R} + b\tilde{R}^2$
discussed in \cite{Chaichian:2010yi} that corresponds to
\[
F(A, B) = F \bigl( (3 - 9\lambda) A^2 + 2\mu B \bigr)
\propto b (3 - 9\lambda)^2 A^4 + 2b\mu(3 - 9\lambda) A^2 B + 4b\mu^2 B^2
+ (3 - 9\lambda) A^2 + 2\mu B \, ,\nonumber
\]
so that we would have
\[
\frac{\p^2 F(A, B)}{\p A^2} \propto 12b (3 - 9\lambda)^2 A^2 + 4b\mu(3 - 9\lambda) B
+ 2(3 - 9\lambda) \, ,\quad
\frac{\p^2 F(A, B)}{\p A \p B} \propto 4b\mu(3 - 9\lambda) A \, ,\quad
\frac{\p^2 F(A, B)}{\p B^2} \propto 8b\mu^2 \, .
\]
}
In this case the equations \eqref{t_constraint1_FRW}--\eqref{t_constraint3_FRW} are
restrictions on the Lagrange multipliers, constituting an
inhomogeneous linear equation for the unknown multipliers
$\lambda_i, i=2,3,4$. Since the homogeneous part
of this equation has only the null solution
$\lambda_2 = \lambda_3 = \lambda_4 = 0$,
the most general solution is the solution of the inhomogeneous equation:
\bea
\label{lambda234_solved}
\lambda_2 &=& N u_2 \equiv - \frac{N}{3} \left( B
 - \frac{9\pi_\beta A}{a^3} \right) \frac{\p^2 F(A, B)}{\p A^2} \nn
&+& \frac{N}{3} \left[ \frac{2}{a}\sum_{i=1}^3 \pi^{ii} + \alpha
+ \left( B - \frac{9\pi_\beta A}{a^3} \right) \frac{\p^2 F(A, B)}{\p A \p B} \right]
\frac{\p^2 F(A, B)}{\p A \p B} \left( \frac{\p^2 F(A, B)}{\p B^2} \right)^{-1} \, ,\nn
\lambda_3 &=& N u_3 \equiv \frac{N}{3} \left( B - \frac{9\pi_\beta A}{a^3} \right) \, ,\nn
\lambda_4 &=& N u_4 \equiv - \frac{N}{3} \left[ \frac{2}{a}\sum_{i=1}^3 \pi^{ii}
+ \alpha + \left( B - \frac{9\pi_\beta A}{a^3} \right) \frac{\p^2 F(A, B)}{\p A \p B} \right]
\left( \frac{\p^2 F(A, B)}{\p B^2} \right)^{-1} \, .
\eea
The multiplier $\lambda_1$ is arbitrary, as is the non-dynamical variable $N$
that also is a multiplier in the Hamiltonian \eqref{H_T_FRW} with \eqref{H_as_constraint}.

The total Hamiltonian \eqref{H_T_FRW} can be written as a sum of two
first-class constraints multiplied by the two arbitrary
time-dependent multipliers $N$ and $\lambda_1$: 
\be 
H_T = N H_0 + \lambda_1 \Phi_1 \, ,
\ee 
where we have defined the first-class Hamiltonian constraint by 
\be \label{H_0_FRW} 
H_0 = \Phi_0 + \sum_{n=2}^4 \int\rd^3 \vect{x}\, u_n(\vect{x})
\Phi_n(\vect{x}) \,, 
\ee 
with the fields $u_n$ ($n=2,3,4$) given by \eqref{lambda234_solved}. 
It is easy to see that $\Phi_1 = \pi_N$ is first-class, since it clearly
has a vanishing Poisson bracket with
every constraint. From \eqref{p_time_evo} and \eqref{s_time_evo} 
we see that the sum of constraints $H_0$ is first-class by 
construction. Note that \eqref{H_0_FRW} is a combination
of secondary and primary constraints. Usually a secondary
first-class constraint would require us to define an extended
Hamiltonian where the constraint would be added with an additional
arbitrary multiplier. In this case, however, that would only lead to
a redefinition of the multiplier $N$, and such a change '$N
\rightarrow N + \text{an arbitrary function of time}$' does not
bring anything new to the description. As always, the first-class
constraints are associated with the gauge symmetries of the system
\cite{Cabo:1993}. The first-class constraints $H_0$ and $\Phi_1$
generate the (gauge) transformations that do not change the physical
state of the system.

The constraints $\chi_k = (\Phi_2, \Phi_3, \Phi_4, \Phi_5, \Phi_6, \Phi_7)$
form the set of second-class constraints of the system.\footnote{The
index of $\chi_k$ runs over $k=1,2,\ldots,6$, so that $\chi_k = \Phi_{k+1}$.}
For details on
the classification and representation of second-class constraints, one can
see \cite{Chaichian:1994}.
The Poisson brackets of the second-class constraints define the matrix:
\be
C_{kl}(\vect{x}, \vect{y}) \equiv \{ \chi_k(\vect{x}), \chi_l(\vect{y}) \}
= C_{kl}(\vect{x}) \delta(\vect{x}-\vect{y}) \, ,
\ee
where
\be
C_{kl}(\vect{x}) = \left(\begin{array}{cccccc}
0 & 0 & 0 & 0 & - 1 & 0 \\
0 & 0 & 0 & - a^3 & - F_{A^2} & - F_{AB} \\
0 & 0 & 0 & 0 & - F_{AB} & - F_{B^2} \\
0 & a^3 & 0 & 0 & 0 & \frac{1}{3} \\
1 & F_{A^2} & F_{AB} & 0 & 0 & 0 \\
0 & F_{AB} &  F_{B^2} & - \frac{1}{3} & 0 & 0
\end{array}\right)
\ee
and we denote
\be
F_{A^2} \equiv \frac{\p^2 F(A, B)}{\p A^2} \, ,\quad F_{AB} \equiv
\frac{\p^2 F(A, B)}{\p A \p B} \, ,\quad F_{B^2} \equiv
\frac{\p^2 F(A, B)}{\p B^2} \, .\label{s_p_derivatives_of_F}
\ee
This matrix has the inverse
\be
C^{kl}(\vect{x}, \vect{y}) = C^{kl}(\vect{x}) \delta(\vect{x}-\vect{y}) \, ,
\ee
\be
\label{inverse_of_C}
C^{kl}(\vect{x}) = \left(\begin{array}{cccccc}
0 & \frac{F_{AB}}{3a^3 F_{B^2}} & - \frac{F_{A^2}}{3a^3 F_{B^2}} &
\frac{F_{AB}^2-F_{A^2}F_{B^2}}{a^3 F_{B^2}} & 1 & - \frac{F_{AB}}{F_{B^2}} \\
 - \frac{F_{AB}}{3a^3 F_{B^2}} & 0 & \frac{1}{3a^3 F_{B^2}}
& \frac{1}{a^3} & 0 & 0 \\
\frac{F_{A^2}}{3a^3 F_{B^2}} & - \frac{1}{3a^3 F_{B^2}} & 0
& - \frac{F_{AB}}{a^3 F_{B^2}} & 0 & \frac{1}{F_{B^2}} \\
\frac{F_{A^2}F_{B^2}-F_{AB}^2}{a^3 F_{B^2}} & - \frac{1}{a^3}
& \frac{F_{AB}}{a^3 F_{B^2}} & 0 & 0 & 0 \\
 - 1 & 0 & 0 & 0 & 0 & 0 \\
\frac{F_{AB}}{F_{B^2}} & 0 & - \frac{1}{F_{B^2}} & 0 & 0 & 0
\end{array}\right)  \, ,
\ee
which satisfies
\be \int \rd^3 \vect{z}\, C_{kl}(\vect{x},
\vect{z}) C^{lm}(\vect{z}, \vect{y}) = C_{kl}(\vect{x})
C^{lm}(\vect{x}) \delta(\vect{x}-\vect{y}) = \delta_k^m
\delta(\vect{x}-\vect{y}) \, .
\ee
Now it is possible to impose the second-class constraints $\chi_k$ by
replacing the Poisson bracket with the Dirac bracket. For any two
functions or functionals $f$ and $h$ of the canonical variables, the
Dirac bracket is defined by
\be \label{DB_FRW}
\{ f(\vect{x}), h(\vect{y}) \}_\mr{DB} = \{
f(\vect{x}), h(\vect{y}) \} - \int \rd^3 \vect{z} \rd^3 \vect{z'} \{
f(\vect{x}), \chi_k(\vect{z}) \} C^{kl}(\vect{z}, \vect{z'})  \{
\chi_l(\vect{z'}), h(\vect{y}) \} \, .
\ee
The Dirac bracket takes fully into account how the second-class 
constraints impose relations between the canonical variables. 
Therefore it enables us to set these constraints to vanish strongly,
$\chi_k(\vect{x}) = 0$.
So we have the identities
\be 
\pi_\alpha = \pi_A = \pi_B = 0 \, ,\quad
A = \frac{\pi_\beta}{3a^3} 
\ee 
and 
\be \label{alpha_and_beta} 
\alpha = - \left. \frac{\p F(A, B)}{\p A} \right|_{A =
\frac{\pi_\beta}{3a^3}} \, ,\quad \beta = - \frac{\p
F(\frac{\pi_\beta}{3a^3}, B)}{\p B} \, .
\ee
When the function $F$ is known, from \eqref{alpha_and_beta} we can
solve the variable $B$
in terms of $\beta$ and $\frac{\pi_\beta}{3a^3} =
\frac{\pi_\beta}{3\sqrt{g}}$:
\be \label{Psi}
B = \tilde{B} \left(\beta, \frac{\pi_\beta}{3a^3}\right)\, .
\ee
Then $\alpha$ can be solved:
\be
\label{alpha}
\alpha = - \left. \frac{\p F\left(A, \tilde{B} \left(\beta,
\frac{\pi_\beta}{3a^3}\right)\right)}{\p A} \right|_{A = \frac{\pi_\beta}{3a^3}} \, .
\ee
Introducing these strong constraints into the Hamiltonian gives
\be
\mc{H} = \frac{2\pi_\beta}{3a} \sum_{i=1}^3 \pi^{ii} - a^3 \left[ \beta\,
\tilde{B} \left(\beta, \frac{\pi_\beta}{3a^3}\right) + F\left(\frac{\pi_\beta}{3a^3},
\tilde{B} \left(\beta, \frac{\pi_\beta}{3a^3}\right)\right) \right] \, .
\ee
The first-class Hamiltonian \eqref{H_0_FRW} reduces to $H_0 = \Phi_0$
 and the total Hamiltonian becomes
\be
\label{H_T_final}
H_T = N\Phi_0 + \lambda_1 \Phi_1 = N \int\rd^3 \vect{x} \mc{H} + \lambda_1 \pi_N \, .
\ee
The canonical variables are $N, \pi_N, \g_{ij}, \pi^{ij}, \beta, \pi_\beta$.
In other words $\alpha, A, B$ and their conjugated momenta have been eliminated.

In order to obtain the equations of motion,
\be
\dot{f} = \{ f, H_T \}_\mr{DB} = N \{ f, \Phi_0 \}_\mr{DB} + \lambda_1 \{ f, \pi_N \}_\mr{DB} \, ,
\ee
for the canonical variables we have to work out all the Dirac brackets
\eqref{DB_FRW} between the variables. We find that the Dirac bracket 
\eqref{DB_FRW} reduces to the Poisson bracket \eqref{PB} for all the 
canonical variables $N, \pi_N, \g_{ij}, \pi^{ij}, \beta, \pi_\beta$, 
and consequently for any functions of these variables.
In the first pair of variables, $N$ is quite arbitrary and $\pi_N$
does not evolve due to the equations of motion:
\bea
 && \dot{N} = \{ N, H_T \}_\mr{DB} = \lambda_1 \{ N, \pi_N \} =
\lambda_1 \, ,\nn 
&& \dot{\pi}_N = \{ \pi_N, H_T \}_\mr{DB} = \{ \pi_N, N \} \int\rd^3
\vect{x} \mc{H} = - \int\rd^3 \vect{x} \mc{H} \approx 0 \, ,
\eea
where as before $\lambda_1$ is an abitrary function of time. For the
spatial metric we get
\be \dg_{ij} = \{ \g_{ij}, H_T \}_\mr{DB} =
\frac{2N\pi_\beta}{3a} \delta_{ij} \, ,
\ee 
where $\dg_{ij}=2a\dot{a}\delta_{ij}$. Solving for $a$ gives
\be
a(t)^3 = a(t_0)^3 + \int_{t_0}^t \rd t N\pi_\beta \, .
\ee 
Hence we need $\pi_\beta$ in order to get $a(t)$. This reveals that
$\pi_\beta$ does not depend on the spatial coordinate $\vect{x}$,
because both $a$ and $N$ depend only on the time coordinate $t$. For
the conjugated momenta we obtain the equations of motion
\be
\dot{\pi}^{ij} = \{ \pi^{ij}, H_T \}_\mr{DB} = \delta^{ij} N \left(
\frac{\pi_\beta}{3a^3} \sum_{k=1}^3 \pi^{kk} + \frac{3a}{2} \left[
\beta \tilde{B} + F(A, \tilde{B}) \right] - \frac{\pi_\beta}{2a^2}
\frac{\p F(A, \tilde{B})}{\p A} \right)_{A=\frac{\pi_\beta}{3a^3}}
\, ,
\ee
where the arguments of $\tilde{B}$ are omitted for brevity,
$\tilde{B} \equiv \tilde{B}(\beta, A) = \tilde{B}\left(\beta,
\frac{\pi_\beta}{3a^3}\right)$, as will be in the next equation. For
the variable $\beta$ we obtain the equation of motion 
\be
\dot{\beta} = \{ \beta, H_T \}_\mr{DB} = \frac{N}{3} \left(
\frac{2}{a} \sum_{i=1}^3 \pi^{ii} - \left. \frac{\p F(A,
\tilde{B})}{\p A} \right|_{A=\frac{\pi_\beta}{3a^3}} \right) \, .
\ee
For its conjugated momentum $\pi_\beta$ we obtain the equation of motion
\be
\dot{\pi}_\beta = \{ \pi_\beta, H_T \}_\mr{DB}
= N a^3 \tilde{B} \left(\beta, \frac{\pi_\beta}{3a^3}\right) \, .
\ee
Further progress in the study of dynamics practically requires one to specify
the form of the function $F$, and then solve \eqref{Psi} from
\eqref{alpha_and_beta}.  We can conclude that when the second
partial derivatives \eqref{s_p_derivatives_of_F} of the function $F$
are non-zero,  the proposed general action \eqref{HLF28} defines
a consistent constrained theory.

Let us then briefly consider the cases when some of the second
partial derivatives \eqref{s_p_derivatives_of_F} of the function $F$
are zero. In such cases the tertiary constraints
\eqref{t_constraint1_FRW}--\eqref{t_constraint3_FRW} are no longer
mere restrictions on the Lagrange multipliers, but in addition
impose constraints on the canonical variables. As an example we
consider the case when $F_{B^2}=0$ and $F_{A^2}\neq 0, F_{AB}\neq
0$. Then we obtain the tertiary constraint 
\be \Phi_8 \equiv
\frac{1}{3} \left( \frac{2}{a}\sum_{i=1}^3 \pi^{ii} + \alpha \right)
+ \left(\frac{B}{3} - \frac{3\pi_\beta A}{a^3}\right) F_{AB} \approx
0 \, , \ee and solve two of the Lagrange multipliers, say
$\lambda_3$ and $\lambda_4$: 
\be 
\lambda_3 = N \left(\frac{B}{3} -
\frac{3\pi_\beta A}{a^3}\right)\, ,\quad \lambda_4 = -
\frac{1}{F_{AB}} \left[ \lambda_2 + N\left(\frac{B}{3} -
\frac{3\pi_\beta A}{a^3}\right) F_{A^2} \right]\, , 
\ee 
where the third multiplier $\lambda_2$ is arbitrary. The consistency
condition
$\dot{\Phi}_8 \approx 0$ of the tertiary constraint $\Phi_8$ imposes
a quartic constraint on the canonical variables, because
$\dot{\Phi}_8$ turns out to be independent of the Lagrange
multiplier $\lambda_2$ and non-vanishing due to the constraints
established so far. Further constraints may follow from the
consistency condition of the quartic constraint. This has to be
checked explicitly after choosing the form of the function $F$.
These additional constraints are a serious threat to the viability
and consistency of the action, since they may delete the physical
degrees of freedom. In case we also have $F_{AB}=0$, we would solve
the Lagrange multipliers $\lambda_2, \lambda_3$ from
\eqref{t_constraint1_FRW}--\eqref{t_constraint3_FRW} and obtain a
tertiary constraint that restricts the field $\beta$ to be a
constant, $\dot{\beta} \approx 0$. Thus in the latter case we should
not have introduced the auxiliary fields $\beta$ and $B$ in the
first place, since $F$ in the action is already linear in its second
argument. We do not discuss the case $F_{A^2}=0$, because it appears
to have very little if any practical application.

As a specific example of the above general theory, one can consider
the FRW cosmology in the modified Ho\v{r}ava-Lifshitz $F(R)$ gravity studied
in \cite{Chaichian:2010yi} and its further generalization considered
in the present work, as action \eqref{HLF11}. However, this analysis
can be used to study FRW cosmology in any theory with an action
of the general form \eqref{HLF26}. Moreover, the methods presented
in this section can be used to analyze any action of the form (\ref{HLF26})
in a general way, without assuming any particular space-time.
The proposed modified Ho\v{r}ava-Lifshitz $F(R)$ gravity will be
studied in the next section.

\section{Hamiltonian analysis of the $F(\tilde{R})$ gravity}
\label{sec:4}

Let us then consider the Hamiltonian analysis of the proposed action
(\ref{HLF11}) for the modified Ho\v{r}ava-Lifshitz $F(R)$ gravity.
The analysis is similar with the Hamiltonian analysis presented in 
ref.~\cite{Chaichian:2010yi}, where a special case of this theory 
given by the choice (\ref{HLFrg0}) was proposed (see also the 
analysis of ref.~\cite{Kluson:2010xx}). This special case with the 
further restriction to the parameter value $\mu=0$ has been proposed
and analyzed in ref.~\cite{Kluson:2009xx}. In this section we generalize 
the analysis of ref.~\cite{Chaichian:2010yi}.

By introducing two auxiliary fields $A$ and $B$ we can write the
action (\ref{HLF11}) into a form that is linear in $\tilde{R}$:
\be
S_{F(\tilde{R})} = \int\rd^4 x \sqrt{\g} N \left[ B(\tilde{R} - A) +
F(A) \right] \, . \label{action_aux}
\ee
Then we can write $\tilde{R}$ as
\be
\tilde R = K_{ij} \mc{G}^{ijkl} K_{kl} + 2\mu \nabla_\mu \left(
n^\mu K \right) - \frac{2\mu}{N} \g[ij] \nabla^{(3)}_i \nabla^{(3)}_j N
 - \mc{L}^{(3)}_R \left(\g_{ij}\right)\ .\label{tildeR_2nd}
\ee
Introducing (\ref{tildeR_2nd}) into (\ref{action_aux}) and
performing integrations by parts yields the action
\bea
S_{F(\tilde{R})} &=& \int\rd t \rd^3 \vect{x} \sqrt{\g}
\Bigl\{ N \left[ B \left( K_{ij} \mc{G}^{ijkl} K_{kl} - \mc{L}^{(3)}_R \left(\g_{ij}\right) - A \right) + F(A) \right] \nn
&&\qquad\qquad\qquad \left.  - 2\mu K \left(  \dot{B} - N^i \p_i B \right)
 - 2\mu N \g[ij] \nabla^{(3)}_i \nabla^{(3)}_j B \right\} \, ,
\label{action_aux_final}
\eea
where the integral is taken over the union $\mc{U}$ of the
$t=\text{constant}$ hypersurfaces $\Sigma_t$ with $t$ over
some interval in $\mathbb{R}$. We assume that the boundary 
integrals on $\p\mc{U}$ and $\p\Sigma_t$ vanish. 
The difference compared to the action studied in ref.~\cite{Chaichian:2010yi} 
is that the potential part $\mc{L}^{(3)}_R(\g_{ij})$ may have any form that 
satisfies the correct scaling property under (\ref{HLF7b}). In other 
words it is not necessarily defined by (\ref{HLFrg0}) and the detailed 
balance condition (\ref{HLF7}). However, due to the projectability 
condition $N = N(t)$ the specific form of the $\mc{L}^{(3)}_R(\g_{ij})$ 
has very little effect on our analysis. Indeed the analysis of 
ref.~\cite{Chaichian:2010yi} is translated to the present more general 
case by making the replacement (\ref{HLFrg0}) from rhs to lhs. 
Therefore we only present the main points of the generalized analysis.

In the Hamiltonian formalism the field variables
$g_{ij}$, $N$, $N^i$, $A$ and $B$ have the canonically conjugated
momenta $\pi^{ij}$, $\pi_N$, $\pi_i$, $\pi_A$ and $\pi_B$,
respectively. For the spatial metric and the field $B$ we have
the momenta
\bea
\pi^{ij} &=& \frac{\delta S_{F(\tilde R)}}{\delta \dot{g}_{ij}}
= \sqrt{\g} \left[ B \mc{G}^{ijkl} K_{kl} - \frac{\mu}{N} \g[ij]
\left(  \dot{B} - N^i \p_i B \right) \right]\, ,\label{metric_momenta}\\
\pi_B &=& \frac{\delta S_{F(\tilde R)}}{\delta \dot{B}}
= - 2\mu \sqrt{\g} K \, .\label{pi_B}
\eea
We assume $\mu\neq 0$ so that the momentum (\ref{pi_B}) does not vanish.
Because the action does not depend on the time derivative of $N$,
$N^i$ or $A$, the rest of the momenta form the set of primary
constraints:
\be
\pi_N \approx 0\, ,\quad \pi_i(\vect{x}) \approx 0\, ,
\quad \pi_A(\vect{x}) \approx 0\, .
\label{p_constraints}
\ee
Because of the projectability condition, the momentum $\pi_N =
\pi_N(t)$ is also constant on $\Sigma_t$ for each $t$. 
Then the Hamiltonian is calculated
\be
H = \int \rd^3 \vect{x} \left( N \mc{H}_0 + N^i \mc{H}_i \right)\, ,\label{Ha}
\ee
where the so-called Hamiltonian constraint and the momentum constraints are 
\bea
\mc{H}_0 &=& \frac{1}{\sqrt{\g}} \left[ \frac{1}{B} \left( \g_{ik} \g_{jl}
 \pi^{ij} \pi^{kl} - \frac{1}{3}\left( \g_{ij} \pi^{ij} \right)^2 \right)
- \frac{1}{3\mu} \g_{ij} \pi^{ij} \pi_B
- \frac{1-3\lambda}{12\mu^2} B \pi_B^2 \right] \nn
&& +\sqrt{\g} \left[ B \left( \mc{L}^{(3)}_R \left(\g_{ij}\right) + A \right)
 - F(A) + 2\mu \g[ij] \nabla^{(3)}_i \nabla^{(3)}_j B \right] \, ,\nn
\mc{H}_i &=& - 2\g_{ij}\nabla^{(3)}_k \pi^{jk} + \nabla^{(3)}_i B \pi_B
\nn
&=& -2\g_{ij}\p_k \pi^{jk} - \left( 2\p_j \g_{ik} - \p_i
\g_{jk} \right) \pi^{jk} + \p_i B \pi_B \, ,\label{Hb}
\eea
respectively. We define the total Hamiltonian by
\be
H_T = H + \lambda_N \pi_N + \int \rd^3 \vect{x} \left( \lambda^i \pi_i
+ \lambda_A \pi_A \right)\, ,
\label{H_T}
\ee
where the primary constraints (\ref{p_constraints}) are multiplied
by the Lagrange multipliers $\lambda_N$, $\lambda^i$, $\lambda_A$.

The primary constraints (\ref{p_constraints}) have to be preserved
under time evolution of the system. Therefore we impose the secondary constraints:
\be
\Phi_0 \equiv \int \rd^3 \vect{x} \mc{H}_0 \approx 0 \, ,\quad
\Phi_i(\vect{x}) \equiv \mc{H}_i(\vect{x}) \approx 0 \, ,\quad
\Phi_A(\vect{x}) \equiv B(\vect{x}) - F'(A(\vect{x}))
\approx 0 \, .\label{s_constraints}
\ee
Here the Hamiltonian constraint $\Phi_0$ is global and the other two,
the momentum constraint $\Phi_i(\vect{x})$ and the constraint
$\Phi_A(\vect{x})$, are local. It is convenient to introduce a
globalised version of the momentum constraints:
\be
\Phi_S(\xi^i) \equiv \int \rd^3 \vect{x}\xi^i \mc{H}_i \approx 0 \, ,
\ee
where $\xi^i, i=1,2,3$ are arbitrary smearing functions. It can be 
shown that the momentum constraints $\Phi_S(\xi^i)$ generate the
spatial diffeomorphisms for the canonical variables $B, \pi_B, \g_{ij},
\pi^{ij}$, and consequently for any function or functional
constructed from these variables, and treats the variables $A,
\pi_A$ as constants.

The consistency of the system requires that also the secondary
constraints $\Phi_0$, $\Phi_S(\xi^i)$ and $\Phi_A(\vect{x})$ have
 to be preserved under time evolution defined by the total Hamiltonian 
(\ref{H_T}), which can be written in terms of the constraints as
\be
H_T = N\Phi_0 + \Phi_S(N^i) + \lambda_N \pi_N
+ \int \rd^3 \vect{x} \left( \lambda^i \pi_i
+ \lambda_A \pi_A \right)\, .
\label{H_T_as_constraints}
\ee
The Poisson brackets for the constraints $\Phi_0$ and $\Phi_S(\xi^i)$ are
\be
\{ \Phi_0, \Phi_0 \} = 0 \, ,\quad \{ \Phi_S(\xi^i), \Phi_0 \} = 0 \, , \quad
\{ \Phi_S(\xi^i), \Phi_S(\eta^i) \} = \Phi_S(\xi^j \p_j \eta^i
 - \eta^j \p_j \xi^i) \approx 0 \, .\label{Phi0_PhiS_PBs}
\ee
For the constraints $\pi_A$ and $\Phi_A(\vect{x})$ the Poisson
brackets that do not vanish strongly are:
\bea
&&\{ \pi_A(\vect{x}), \Phi_0 \} = - \sqrt{\g} \Phi_A(\vect{x}) \approx 0 
\, ,\quad \{ \pi_A(\vect{x}), \Phi_A(\vect{y}) \} = F''(A(\vect{x}))
\delta(\vect{x}-\vect{y})\,,\nn
&&\{ \Phi_0, \Phi_A(\vect{x}) \} = \frac{1}{3\mu\sqrt{\g}} 
\left(\g_{ij} \pi^{ij} + \frac{1-3\lambda}{2\mu}B\pi_B \right) \, , \quad
\{ \Phi_S(\xi^i), \Phi_A(\vect{x}) \} = - \xi^i \p_i B \, .
\label{piA_PhiA_PBs}
\eea
Since $F''(A)=0$ would essentially reproduce the original projectable
Ho\v{r}ava-Lifshitz gravity, we assume that $F''(A) \neq 0$.
The constraint $\Phi_A(\vect{x})$ can be made consistent by
fixing the Lagrange multiplier $\lambda_A$:
\be
\lambda_A = \frac{1}{F''(A)} \left( N^i \p_i B - \frac{N}{3\mu\sqrt{\g}}
\left(\g_{ij} \pi^{ij} + \frac{1-3\lambda}{2\mu}B\pi_B \right) \right)\, .\label{lambda_A}
\ee
Now all the constraints of the system are consistent under dynamics.

According to the Poisson brackets
(\ref{Phi0_PhiS_PBs})--(\ref{piA_PhiA_PBs}) between the constraints,
we can set the second-class constraints $\pi_A(\vect{x})$ and
$\Phi_A(\vect{x})$ to vanish strongly, and as a result turn the
Hamiltonian constraint $\Phi_0$ and the momentum constraint
$\Phi_S(\xi^i)$ into first-class constraints, by replacing the Poisson 
bracket with the Dirac bracket. It turns out that the the Dirac bracket 
reduces to the Poisson bracket for any functions of the canonical 
variables. Assuming we can solve the constraint $\Phi_A(\vect{x})=0$, 
i.e. $B=F'(A)$, for $A=\tilde{A}(B)$, where $\tilde{A}$ is the inverse 
of the function $F'$, we can eliminate the variables $A$ and $\pi_A$. 
Thus the final variables of the system are $\g_{ij}, \pi^{ij}, B, \pi_B$. 
The lapse $N$ and the shift vector $N^i$, together with $\lambda_N$ 
and $\lambda^i$, are non-dynamical multipliers. Finally the total 
Hamiltonian is the sum of the first-class constraints
\be
H_T = N\Phi_0 + \Phi_S(N^i) + \lambda_N \pi_N + \int\rd^3 \vect{x}
\lambda^i \pi_i \, .\label{H_T_sum_of_first-class}
\ee

We conclude that the proposed action (\ref{HLF11}) of the more 
general modified Ho\v{r}ava-Lifshitz $F(R)$ gravity also defines a 
consistent constrained theory when the projectability condition is 
postulated. For additional details and  discussion on the analysis 
see ref.~\cite{Chaichian:2010yi}.

\section{Hamiltonian analysis of the $F(\tilde{R})$ gravity in fixed gauge}
\label{sec:5}

Let us then analyze the action \eqref{HLF11} when the gauge is fixed
by \eqref{HLFrg2}, and we obtain the action \eqref{HLFrg1b}. First we find
the momenta canonically conjugated to the generalized coordinates $\bg_{ij}$
and $\varphi$ of the action \eqref{HLFrg1b}. For the fields $\varphi$
and $\bg_{ij}$ we find the momenta
\be
\label{pi_varphi}
\pi_\varphi = \frac{\delta S_{F(\tilde{R})}}{\delta \dot{\varphi}}
= \frac{\sqrt{\bg}}{4\kappa^2} \left( - (1-3\lambda+3\mu) \bg[ij] \dbg_{ij}
+ 3 (1-3\lambda+6\mu) \dot{\varphi} \right)
\ee
and
\be
\label{bpi}
\bpi^{ij} = \frac{\delta S_{F(\tilde{R})}}{\delta \dbg_{ij}} = \frac{\sqrt{\bg}}{4\kappa^2}
\left( \bg[ik] \bg[jl] \dbg_{kl} - \lambda \bg[ij] \bg[kl] \dbg_{kl}
 - (1-3\lambda+3\mu) \bg[ij] \dot{\varphi} \right)\,,
\ee
respectively.

In the following analysis we will first assume
\be 1-3\lambda+3\mu \neq 0
\, ,\quad 1-3\lambda+6\mu \neq 0 \, ,\quad \mu \neq 0 \, ,
\ee
so that the kinetic term for $\varphi$ does not vanish. Later we will
consider the special cases where these conditions do not hold. First
we solve $\dot{\varphi}$ from \eqref{pi_varphi}, 
\be
\label{dotvarphi} \dot{\varphi} = \frac{\frac{4\kappa^2}{\sqrt{\bg}}
\pi_\varphi + (1-3\lambda+3\mu) \bg[ij] \dbg_{ij}}{3
(1-3\lambda+6\mu)}\,, 
\ee 
and introduce it into \eqref{bpi} 
\be
\label{bpi2} \bpi^{ij} = \frac{\sqrt{\bg}}{4\kappa^2} \left[ \bg[ik]
\bg[jl] \dbg_{kl}
 - \left(\frac{1}{3} + \frac{3\mu^2}{1-3\lambda+6\mu}\right) \bg[ij] \bg[kl]
\dbg_{kl} \right] - \frac{1-3\lambda+3\mu}{3 (1-3\lambda+6\mu)} \bg[ij] \pi_\varphi \, .
\ee
Then we find the velocities in terms of the coordinates and momenta.
First we contract \eqref{bpi2} by $\bg_{ij}$ and solve for
\be
\bg[ij] \dbg_{ij} = - \frac{4\kappa^2}{\sqrt{\bg}} \left( \frac{1-3\lambda+6\mu}{9\mu^2}
\bg_{ij} \bpi^{ij} + \frac{1-3\lambda+3\mu}{9\mu^2} \pi_\varphi \right) \, .
\ee
This is inserted back into \eqref{bpi2} as well as into \eqref{dotvarphi},
which enables us to obtain the velocities in terms of the canonical variables:
\begin{gather}
\label{varphi_eom}
\dot{\varphi} = \frac{4\kappa^2}{\sqrt{\bg}} \left( \frac{3\lambda-1}{27\mu^2} \pi_\varphi
 - \frac{1-3\lambda+3\mu}{27\mu^2} \bg_{ij} \bpi^{ij} \right) \, ,\\
\label{bg_eom}
\dbg_{ij} = \frac{4\kappa^2}{\sqrt{\bg}} \left\{ \bg_{ik} \bg_{jl} \bpi^{kl}
 - \bg_{ij} \left[ \left(\frac{1-3\lambda+6\mu}{27\mu^2}+\frac{1}{3}\right)
\bg_{kl} \bpi^{kl} + \frac{1-3\lambda+3\mu}{27\mu^2} \pi_\varphi \right] \right\} \, .
\end{gather}
Thus there are no primary constraints. It is expected that there are no first-class constraints,
since the gauge has been completely fixed by setting $N=1, N^i=0$.
Due to the non-vanishing kinetic terms of $\varphi$, there are no second-class constraints either.
The Hamiltonian is defined by
\be
H = \int\rd^3 \vect{x} \left( \bpi^{ij} \dbg_{ij} + \pi_\varphi \dot{\varphi} \right) - L \, .
\ee
After a lengthy algebra exercise we find
\bea
\label{H_fixed_gauge}
H &=& \int\rd^3 \vect{x} \Biggl\{ \frac{4\kappa^2}{\sqrt{\bg}} \biggl[ \frac{1}{2}
\bg_{ik} \bg_{jl} \bpi^{ij} \bpi^{kl} - \left(\frac{1-3\lambda+6\mu}{54\mu^2}
+\frac{1}{6}\right) \left( \bg_{ij} \bpi^{ij} \right)^2 \nn
&&\qquad\quad - \frac{1-3\lambda+3\mu}{27\mu^2} \bg_{ij} \bpi^{ij} \pi_\varphi
+ \frac{3\lambda-1}{54\mu^2} \pi_\varphi^2 \biggr] - \frac{\sqrt{\bg}}{2\kappa^2}
\left[ \bcLR \left(\bg_{ij}, \varphi\right) - V(\varphi) \right] \Biggr\} \, .
\eea
In fact we find that the Hamiltonian \eqref{H_fixed_gauge} is correct
for any parameters $\lambda$ and $\mu$ as long as it is defined, i.e. 
when $\mu \neq 0$. This can be seen by considering the two cases when the 
kinetic cross-term vanishes ($\dot{\varphi}$ and $\dbg_{ij}$ decouple),
$1-3\lambda+3\mu = 0, \mu \neq 0$, and when the kinetic 
$\dot{\varphi}^2$ term vanishes, $1-3\lambda+6\mu = 0, \mu \neq 0$,
separately.  Even the formulas \eqref{varphi_eom}--\eqref{bg_eom} 
hold in these cases, thought the details of their calculation are 
quite diferent. Note that the vanishing of the both kinetic terms of 
$\varphi$ implies $\mu = 0$ and $\lambda=1/3$.

The Poisson bracket is postulated by (equal time $t$ is understood)
\be
\{ \bg_{ij}(\vect{x}), \bpi^{kl}(\vect{y}) \} = \frac{1}{2} \left( \delta_i^k \delta_j^l
+ \delta_i^l \delta_j^k \right) \delta(\vect{x} - \vect{y})\, ,\quad \{ \varphi(\vect{x}),
\pi_\varphi(\vect{y}) \} = \delta(\vect{x} - \vect{y}) \, ,
\ee
with all the other Poisson brackets vanishing. Now we can work out
the Hamiltonian equations of motion. Because there are no constraints, 
the Hamiltonian \eqref{H_fixed_gauge} defines the dynamics of any 
function or functional $f$ of the variables
$\bg_{ij}, \bpi^{ij}, \varphi, \pi_\varphi$ by:
\be
\dot{f} = \{ f, H \} \, .
\ee
For the generalized coordinates $\bg_{ij}$ and $\varphi$ we obtain the equations
of motion \eqref{bg_eom} and \eqref{varphi_eom} respectively.
For the momenta $\bpi^{ij}$ we obtain
\bea
\label{dot_bpi}
\dot{\bpi}^{ij} &=& \frac{4\kappa^2}{\sqrt{\bg}} \biggl[ \bg[ij]
\left( \frac{1}{4} \bg_{km} \bg_{ln} \bpi^{kl} \bpi^{mn} - \frac{a}{4}
\left( \bg_{kl} \bpi^{kl} \right)^2 - \frac{b}{2} \bg_{kl} \bpi^{kl} \pi_\varphi
+ \frac{c}{4} \pi_\varphi^2 \right)\nn
&&\qquad\qquad - \bg_{kl} \bpi^{ik} \bpi^{jl} + a \bpi^{ij} \bg_{kl} \bpi^{kl} 
+ b \bpi^{ij} \pi_\varphi \biggr] \nn
&&+ \frac{\sqrt{\bg}}{2\kappa^2} \left( \frac{1}{2} \bg[ij]
\left[ \bcLR \left(\bg_{ij}, \varphi\right) - V(\varphi) \right]
+ \frac{\p \bcLR \left(\bg_{kl}, \varphi\right)}{\p \bg_{ij}} \right) \, ,
\eea
where we have introduced the constants:
\be
a = \frac{1-3\lambda+6\mu}{27\mu^2}+\frac{1}{3} \, ,\quad
b = \frac{1-3\lambda+3\mu}{27\mu^2} \, ,\quad c = \frac{3\lambda-1}{27\mu^2} \, .
\ee
The equation of motion for $\pi_\varphi$ is
\be
\dot{\pi}_\varphi = \frac{\sqrt{\bg}}{2\kappa^2}
\left( \frac{\p \bcLR \left(\bg_{kl}, \varphi\right)}{\p \varphi}
 - 3A(\varphi) \e^{3\varphi} \right) \, ,
\ee
where the last term is obtained from the derivative of $V(\varphi)$
from \eqref{HLFrg5}
\be
\frac{\rd V(\varphi)}{\rd\varphi}
= A(\varphi) \frac{\rd A(\varphi)}{\rd\varphi} F''(A(\varphi))
= 3A(\varphi) F'(A(\varphi)) = 3A(\varphi) \e^{3\varphi} \, .
\ee
Here we have also used the definition \eqref{varphi} of $\varphi$
and \eqref{Avarphi} of $A(\varphi)$ to calculate
\be
1 = \frac{\rd \varphi}{\rd\varphi} = \frac{1}{3} \frac{\frac{\rd A(\varphi)}{\rd\varphi}
F''(A(\varphi))}{F'(A(\varphi))} \
\Rightarrow\  \frac{\rd A(\varphi)}{\rd\varphi} F''(A(\varphi)) = 3 F'(A(\varphi)) \, .
\ee
In particular, the equations of motion \eqref{dot_bpi} for the momenta $\bpi^{ij}$
are pretty complex. However, as always, they are first order differential equations.

The cases with $\mu=0$ are less interesting and we only consider them
briefly. When $\lambda=1/3$ the field $\varphi$ is non-dynamical and
hence one has the primary constraint $\pi_\varphi \approx 0$. The momentum conjugate to $\bg_{ij}$ are given by
\be
\bpi^{ij} = \frac{\sqrt{\bg}}{4\kappa^2} \left( \bg[ik] \bg[jl] \dbg_{kl} - \frac{1}{3}\bg[ij] \bg[kl] \dbg_{kl} \right) \, .\label{bpi_traceless}
\ee
It has zero trace $\bg_{ij}\bpi^{ij}=0$ and can be trivially solved for
$\dbg_{ij}=\frac{4\kappa^2}{\sqrt{\bg}}\bg_{ik}\bg_{jl}\bpi^{kl}$.
When $\lambda\neq 1/3$, one is forced to impose the constraint
$\pi_\beta=-\bg_{ij}\bpi^{ij} \approx 0$ that again leads to
(\ref{bpi_traceless}) and makes $\varphi$ non-dynamical.

\section{FRW cosmology in power-like models: cosmic acceleration and future singularities}
\label{sec:6}

We now consider the FRW cosmology of the action (\ref{HLF11}). 
In the spatially-flat FRW space-time (\ref{HLF8}), since the spatial 
curvature vanishes, $R^{(3)}_{ij}=R^{(3)}=0$, there is no contribution 
from $\mathcal{L}_R^{(3)}$, as it vanishes according to (\ref{HLFrg9}) 
or (\ref{HLFrg14}). In other words, the choice of $\mathcal{L}_R^{(3)}$
in (\ref{HLFrg9}) or (\ref{HLFrg14}) gives the same FRW cosmology.
Of course, this situation changes when one considers black holes or
other solutions with non-trivial dependence on the spatial
coordinates. 

Let us first review the spatially-flat FRW equations obtained in ref. 
\cite{Chaichian:2010yi}. Varying the action (\ref{HLF11}) with 
respect to $g^{(3)}_{ij}$ and setting $N=1$ one obtains:
\be \label{HLF13} 0 = F\left(\tilde
R\right) - 2 \left(1 - 3\lambda + 3\mu \right) \left(\dot H + 3
H^2\right) F'\left(\tilde R\right) - 2\left(1 - 3\lambda \right) H
\frac{\rd F'\left(\tilde R\right)}{\rd t} + 2\mu \frac{\rd^2
F'\left(\tilde R\right)}{\rd t^2} + p\, ,
\ee
where $F'$ denotes the
derivative of $F$ with respect to its argument. Here, the matter
contribution (the pressure $p$) is included. On the other hand, the
variation over $N$ gives the global constraint: 
\be \label{HLF14}
0 = \int \rd^3 \bm{x} \left[ F\left(\tilde R\right) - 6 \left\{ \left(1 -
3\lambda + 3\mu\right) H^2 + \mu \dot H \right\} F'\left(\tilde
R\right) + 6 \mu H \frac{\rd F'\left(\tilde R\right)}{\rd t} - \rho
\right]\, ,
\ee
after setting $N=1$. Here $\rho$ is the energy
density of matter and we have set again $N=1$. It is important to
stress that, because of the projectability condition $N=N(t)$, the
above equation is a global constraint. If the standard conservation
law is used, 
\be \label{HLF15} 
0= \dot \rho + 3H \left(\rho + p\right)\, , 
\ee 
Eq. (\ref{HLF13}) can be integrated to
give:\footnote{Note that, as already shown in \cite{Carloni:2009jc}
for the standard case, the  parameter $\lambda$ has a crucial role
in the relation of Ho\v{r}ava-Lifshitz type theories. In fact, from
the second equation above one realizes that this solution is
physical only if  $1-3 \lambda +3\mu>0$. It should also be pointed
out that in Ho\v{r}ava-Lifshitz gravity the role of standard matter
and its conservation properties are not well understood yet. We will
proceed with our discussion supposing that it is possible to couple
matter and gravity in the same way in which one does in GR.} 
\be
\label{HLF16} 0 = F\left(\tilde R\right) - 6 \left\{ \left(1 -
3\lambda + 3\mu\right) H^2 + \mu \dot H \right\} F'\left(\tilde
R\right) + 6 \mu H \frac{\rd F'\left(\tilde R\right)}{\rd t} - \rho
- \frac{C}{a^3}\, . 
\ee 
Here $C$ is the integration constant and can
be set to zero. In \cite{Mukohyama:2009mz}, however, it has been
claimed that $C$ does not necessarily need to vanish in a local
region, since (\ref{HLF14}) needs to be satisfied only in the whole
universe. In this sense in a limited region, one can have $C>0$ and
the $Ca^{-3}$ term in (\ref{HLF16}) can be regarded as dark matter.

Note that Eq. (\ref{HLF16}) corresponds to the first FRW equation
and (\ref{HLF13}) to the second one. Specifically, if we choose
$\lambda=\mu=1$ and $C=0$, Eq. (\ref{HLF16}) reduces to
\bea
\label{HLF17}
0 &=& F\left(\tilde R\right)
- 6 \left(H^2 + \dot H \right) F'\left(\tilde R\right)
+ 6 H \frac{\rd F'\left(\tilde R\right)}{\rd t} - \rho \nn
&=& F\left(\tilde R\right) - 6 \left(H^2 + \dot H \right)
F'\left(\tilde R\right) + 36 \left(4H^2 \dot H + \ddot H\right)
F''\left(\tilde R\right) - \rho \, , \eea which is identical to the
corresponding equation in the standard $F(R)$ gravity (see Eq. (2)
in \cite{Nojiri:2009kx} where a reconstruction of the theory has
been made). In the following we will explore the properties of the
equations  (\ref{HLF13}) and (\ref{HLF16}), especially looking for
solutions that represent accelerated expansion. Solutions of this
type are very important because they represent the key evolutionary
phases of the universe, namely the inflationary era and the dark
energy era. The connection with dark energy is particularly
important to understand, if the Newtonian nature of the quantum
theory of gravitation, implicit in Ho\v{r}ava-Lifschitz gravity, is
the direct cause of cosmic acceleration and, as a consequence, of
dark energy.

\subsection{de Sitter cosmology}

Let us investigate the properties of the de Sitter solutions in this
class of theories. This issue was considered for the first time in 
\cite{Chaichian:2010yi}, but in the following a more general treatment 
is given. These solutions are of great importance in
cosmology because they have the potential to describe both
inflationary phase(s) as well as dark energy era(s). In standard
$F(R)$ gravity it has been proven that it is possible to construct a
viable model unifying inflation  and late time acceleration in the
form of double or multiple de Sitter solution \cite{NO2003,Nojiri:2007as,Cognola:2007zu,Cognola:2008zp}.

In vacuum ($\rho_m, p_m=0$) and substituting the de Sitter metric
\begin{equation} \label{dSmetric}
\rd s^2= -\rd t^2+\exp{(\gamma t)}\sum_{i=1}^{3}\left(\rd x^i\right)^2\,,
\end{equation}
the  equations (\ref{HLF13}) and (\ref{HLF16}) reduce to the single equation
\begin{equation} \label{EqdS}
0=F\left(\tilde R\right)+6 \gamma ^2 \left(3 \lambda -3 \mu -1\right) F'\left(\tilde R\right)\,.
\end{equation}
In Table \ref{dSTab} we show the values of the time constant $\gamma$ of
the de Sitter metric of some popular $F(R)$ theories and their
Ho\v{r}ava-Lifshitz versions.

It is interesting to note that, in contrast to what happens in
standard $F(R)$ gravity, the quadratic function 
$F(\tilde R)=\tilde{R}^m$ is
not degenerate for $m=2$, but only for
\begin{equation}
m=\frac{2}{1-3 \lambda +3 \mu }\,,
\end{equation}
i.e. it depends on the Lorentz-violation parameters.

It is also interesting to note that in general the solution of
(\ref{HLF13}) and (\ref{HLF16}) is not unique. Thus a given theory
can have multiple de Sitter solutions.  This means that also in this
case the cosmologies of these theories can admit both  inflation and
dark energy phases. However, since the  Ho\v{r}ava-Lifshitz
parameters are in principle only present in the coefficients of the
equation (\ref{EqdS}) and the number of solutions of (\ref{EqdS}) is
determined by the powers of $\tilde{R}$ appearing in $F$, two
corresponding theories will have in general the same number of de
Sitter solutions.  Obvious exceptions are  the case $3 \lambda -3
\mu -1=0$ for which the Eq. (\ref{EqdS})   becomes $F=0$  and the
case $3 \lambda -3 \mu -1=g(\gamma)$.

In the first case, we see that the structure of the cosmological
equations is essentially changed. In particular, the equation
(\ref{HLF13}) is modified and the constraint
 (\ref{HLF16}) looses the linear $H^2$ term:
\be \label{HLFrg15} 0 = F\left(\tilde R\right) - 6 \mu \dot H
F'\left(\tilde R\right) + 6 \mu H \frac{\rd F'\left(\tilde
R\right)}{\rd t} - \rho\, \ee (we have considered $C=0$).
Consequently, the equation (\ref{EqdS}) becomes 
\be \label{HLFrg16}
0 = F\left(\tilde R\right)\, , 
\ee 
which is never obtained in the standard $F(R)$ gravity.

In the second case, instead, the choice of the function $g$ can
radically change the number and type of solutions in these theories.
In this sense, the solution space for Ho\v{r}ava-Lifshitz $F(R)$
gravity can be considered bigger than  the one of its standard
counterpart. Such a fact will be even more apparent in the case of
the FRW-type solutions that will be examined in the next section.

\begin{table}[htdp]
\caption{Some of the values of the time constant of  de Sitter backgrounds
for standard $F(R)$ gravity models and their Ho\v{r}ava-Lifshitz counterparts.
When writing the form of the function $F$, the Ricci scalar of both types of
theories is indicated by $x$. For the more complex forms of $F(R)$ an implicit
equation has to be solved for the time parameter $\gamma$ in order to find its
values.}\label{dSTab}
\begin{center}
\scriptsize{\begin{tabular}{cccc}
\hline\hline
Function $F$& $ \mbox{Standard Case} $ & $\mbox{Ho\v{r}ava-Lifshitz case}$
\\\hline
$x+\chi x^n$ &  $\gamma=\pm\left(2^{2 n-1} 3^{n-1} \alpha -3^n 4^{n-1} n \chi
\right)^{\frac{1}{2-2 n}}$ & $\gamma=\pm\left(\frac{2^{2 n} 3^n \chi  \left(3
n \lambda -3 n \mu -n+2\right)}{-36 \lambda +36 \mu -12}\right)^{\frac{1}{2
(1-n)}}$
\\&&&\\
$x^n\exp(\chi x^m)$ &  $\gamma=\frac{1}{2 \sqrt{3}}\left(\frac{2-n}{m
\chi}\right)^{\frac{1}{2 m}}$ & $\gamma=\frac{1}{2 \sqrt{3}}\left[\frac{n
\left(3 \lambda -3 \mu
 -1\right)+2}{ \chi m \left(3 \mu +1-3 \lambda \right)} \right]^{\frac{1}{2
m}}$
\\&&&\\
$\frac{x^m+\chi}{1+\xi x^n}$ &
$\frac{A}{2 \left(12^n \gamma ^{2 n}+\xi \right)^2}=0$ &$\frac{B}{2
\left(12^n \gamma ^{2 n}+\xi \right)^2}=0$
\\&&&\\
$x+\chi+\frac{\chi }{\alpha  \left[(x \xi -1)^{2
 n+1}+1\right]+1}$& $-C+6 \gamma ^2+\chi=0$& $-3 D+6 \gamma ^2 \left(3
\lambda -3 \mu +1\right)+\chi=0$
\\&&&\\
\hline
\multicolumn4c{$A=2^{2 (m+n)} 3^{m+n} (-m+n+2) \gamma ^{2 (m+n)}+2^{2 m} 3^m
(2-m) \xi  \gamma ^{2 m}+2^{2 n} 3^n (n+2) \chi  \gamma ^{2 n}+2 \xi
\chi$}\\\multicolumn4c{$B=2^{2 (m+n)} 3^{m+n} \gamma ^{2 (m+n)} \left(m
\left(3 \lambda -3 \mu -1\right)-3 n \lambda +3 n \mu
+n+2\right)+~~~~~~~~~~~~~~~~~~~~~~~~~$}\\ \multicolumn4c{$+2^{2 m} 3^m \xi
\gamma ^{2 m} \left(m
 \left(3 \lambda -3 \mu -1\right)+2\right)+2^{2 n} 3^n \chi  \gamma ^{2 n}
\left(-3 n \lambda +3 n \mu +n+2\right)+2 \xi  \chi$}\\
\multicolumn4c{$C=\frac{\chi  \left(-6 (-2 n-3) \gamma ^2 \xi -1\right)-6 (2
n+1) (\alpha +1) \gamma ^2 \xi  \chi}{\left(12 \gamma ^2 \xi -1\right)
\left(\alpha  \left(12 \gamma ^2 \xi -1\right)^{2 n+1}+\alpha
+1\right)^2}$}\\
\multicolumn4c{$D=\frac{6 (2 n+1) (\alpha +1) \gamma ^2 \xi  \chi  \left(3
\lambda -3 \mu -1\right)+\chi  \left(-6 \gamma ^2 \xi  \left(n \left(6
\lambda -6 \mu -2\right)+3
 \left(\lambda -\mu -1\right)\right)-1\right)}{\left(12 \gamma ^2 \xi
-1\right) \left(\alpha  \left(12 \gamma ^2 \xi -1\right)^{2 n+1}+\alpha
+1\right)^2}$}\\\hline\hline
\end{tabular}}
\end{center}
\end{table}

\subsection{Power law solutions and reconstruction
technique}\label{SectionReconstruction}

In addition to the de Sitter solution described above one can also
look for accelerated expansion phases in the form of  power law
solutions. These solution can have a double value as Friedmannian
cosmologies, if the exponent of the power law is in the interval
$]0,1[$, and they can realize the so-called ``power law inflation",
or a ``power law dark energy", if the exponent is bigger than one.

If we look for the presence of Friedmann solutions of (\ref{HLF13}) and
(\ref{HLF16}), we realize quickly that, as in $F(R)$ gravity, there is little
chance to find power law solutions, unless one considers a function $F$
of trivial form. However, due to the additional parameters, the set of solutions
of this type is bigger in the Ho\v{r}ava-Lifshitz case than in the standard one.
For example, in the simple case $F(\tilde{R})=\tilde{R}+\chi \tilde{R}^m$ we
find that, in the presence of a barotropic fluid ($p=w\rho$), the spatially-flat solution
\begin{equation} \label{}
a=a_0 t^{2/3(1+w)}\,, \qquad \rho=\rho_0 t^{-2}\,,
\end{equation}
satisfies (\ref{HLF13}) and (\ref{HLF16}) if
\begin{equation}
\mu =\frac{\left(w^2-1\right) \gamma  \left(3 \lambda -1\right)}{2 w (3 (w+1)
\gamma -w+1)}\,, \quad \mbox{and} \quad \rho _0= \frac{4 \left(3 \lambda
-1\right)}{3 (w+1)^2 \kappa ^2}\,.
\end{equation}
This corresponds to the standard Friedmann solution.
It is well known that in  standard $F(R)$ theories, the case $F(R)=R+\chi
R^m$  possesses only power law solutions of the type $a=a_0 t$  or $a=a_0
t^{1/2}$  (see e.g. \cite{Carloni:2007br}).

In order to facilitate the analysis in the next sections,  we
consider also some solutions for this model in the case of very
small and very large scalar curvature. In the first case, the theory
reduces itself to GR plus a cosmological constant and its solutions
are approximated by the Friedmann ones. In the case of high
curvature, instead, the theory reduces to
$F(\tilde{R})\approx\tilde{R}^n$. Such a theory possesses three
exact solutions. The first two
\begin{eqnarray} \label{SolMat}
&&a=a_0 t^{\gamma}\,, \qquad \rho=\rho_0 t^{-2}\,,\qquad
\gamma=\frac{2 m}{3 (1 + w)}\,,\\ &&\rho_0 =\frac{\chi  \left[3
\mu(m-1)  (2 m (w+2)-w-1)-m (2 m-1) \left(3 \lambda -1\right)\right]
}{\kappa ^2 \left(m \left[3 \lambda -6 \mu -1\right)+3 (w+1) \mu
\right]}\left(\frac{4 m^2 \left(-3 \lambda +6 \mu +1\right)-12 m
(w+1)\mu }{(w+1)^2}\right)^m\,,\nonumber
\end{eqnarray}
and
\begin{eqnarray} \label{SolVac}
&&a=a_0 t^{\gamma}\,, \qquad \rho=\rho_0 t^{-2}\,,\qquad \gamma=\frac{2 (m-1)
(2 m-1) \mu }{(2 m-1) \left(3 \lambda -1\right)-6 (m-1) \mu }\,,\qquad
\rho_0 =0\,,
\end{eqnarray}
correspond to the solutions in the standard $F(R)$ case.
A third solution, that is valid only for $m>1$, is
\begin{eqnarray} \label{Sol HL}
&&a=a_0 t^{\gamma}\,, \qquad \rho=\rho_0 t^{-2}\,,\qquad \gamma=\frac{2 \mu
}{-3 \lambda +6 \mu +1}\,,\qquad \rho_0 =0\,,
\end{eqnarray}
which is characteristic of Ho\v{r}ava-Lifshitz gravity and does not depend on $m$. In the
analysis of the singularities of this simple model, we will refer to these
solutions.

Note that as it often happens \cite{Carloni:2004kp} in theories of this type,
the value of $\rho_0$ can be negative (or even undefined)
for certain combinations of variables. This implies that matter is not
always compatible with $F(R)$ gravity, not even in the Ho\v{r}ava-Lifshitz case.
In our specific example $\rho_0>0$ implies
\begin{eqnarray}
&&m<0,\qquad 0\leq w\leq 1, \quad\left\{
\begin{array}{ccc}
\chi <0 & \mu >0 &\frac{6 m \mu +m-3 w \mu -3 \mu }{3 m}<\lambda <\frac{6
m^2 w \mu +12 m^2 \mu +2 m^2-9 m w\mu -15 m \mu -m+3 w \mu +3 \mu }{6 m^2-3
m}   \\
\chi >0&\mu <0 &\lambda >\frac{6 m \mu +m-3 w \mu -3 \mu }{3
 m}  \\
\chi >0&\mu \geq 0 & \lambda >\frac{6 m^2 w \mu +12 m^2 \mu +2 m^2-9 m w \mu
-15 m \mu -m+3 w \mu +3 \mu }{6 m^2-3m}
\end{array}
\right.
\\
&&\chi >0, \qquad \mu >0 , \quad\left\{
\begin{array}{ccc}
w=0&0<m<\frac{1}{2} &\lambda <\frac{6 m \mu +m-3 \mu }{3 m}   \\
w=0&m>\frac{1}{2} &\frac{12 m^2 \mu +2 m^2
-15 m \mu -m +3 \mu }{6 m^2-3m}<\lambda <\frac{6 m \mu +m-3 \mu }{3 m}   \\
0<w\leq 1&0<m< \frac{1}{2} &\lambda <\frac{6 m \mu +m-3 w \mu -3 \mu }{3
 m}  \\
0<w\leq 1&\frac{1}{2}<m< \frac{1+w}{2 w} &\frac{6 m^2 w \mu +12 m^2 \mu +2 m^2-9 m w \mu
-15 m \mu -m+3 w \mu +3 \mu }{6 m^2-3m}<\lambda <\frac{6 m \mu +m-3 w \mu -3 \mu }{3
 m} 
 \end{array}
\right.
\\
&&\chi <0, \qquad \mu >0 , \quad\left\{
\begin{array}{ccc}
w=0&m>\frac{1}{2} &\lambda <\frac{12 m^2 \mu +2 m^2
-15 m \mu -m +3 \mu }{6 m^2-3m} \\
0<w\leq 1&\frac{1}{2}<m< \frac{1+w}{2 w} &\lambda <\frac{6 m^2 w \mu +12 m^2 \mu +2 m^2-9 m w \mu
-15 m \mu -m+3 w \mu +3 \mu }{6 m^2-3m}\\
0<w\leq 1&m>\frac{1+w}{2 w} &\lambda <\frac{6 m \mu +m-3 w \mu -3 \mu }{3
 m} 
 \end{array}
\right.
\\
&&\chi >0, \qquad \mu <0 , \quad\left\{
\begin{array}{ccc}
w=0&m>\frac{1}{2} &\lambda <\frac{12 m^2 \mu +2 m^2
-15 m \mu -m +3 \mu }{6 m^2-3m} \\
0<w\leq 1&\frac{1}{2}<m< \frac{1+w}{2 w} &\lambda <\frac{6 m^2 w \mu +12 m^2 \mu +2 m^2-9 m w \mu
-15 m \mu -m+3 w \mu +3 \mu }{6 m^2-3m}\\
0<w\leq 1&m>\frac{1+w}{2 w} &\frac{6 m^2 w \mu +12 m^2 \mu +2 m^2-9 m w \mu
-15 m \mu -m+3 w \mu +3 \mu }{6 m^2-3m}<\lambda <\frac{6 m \mu +m-3 w \mu -3 \mu }{3
 m} 
 \end{array}
\right.\\
&&\chi <0, \qquad \mu <0 , \quad\left\{
\begin{array}{ccc}
w=0&0<m<\frac{1}{2} &\frac{12 m^2 \mu +2 m^2
-15 m \mu -m +3 \mu }{6 m^2-3m}<\lambda <\frac{6 m \mu +m-3 \mu }{3 m}  \\
w=0&m>\frac{1}{2}&\lambda <\frac{6 m \mu +m-3 \mu }{3 m}     \\
0<w\leq 1&0<m< \frac{1}{2} &\frac{6 m^2 w \mu +12 m^2 \mu +2 m^2-9 m w \mu
-15 m \mu -m+3 w \mu +3 \mu }{6 m^2-3m}<\lambda <\frac{6 m \mu +m-3 w \mu -3 \mu }{3
 m} 
 \\
0<w\leq 1&\frac{1}{2}<m< \frac{1+w}{2 w} &\lambda <\frac{6 m \mu +m-3 w \mu -3 \mu }{3
 m}  \end{array}
\right.
\end{eqnarray}
Another result which will be useful for our purposes is an exact solution
for the theory $F(\tilde{R})=\tilde{R}+\xi \tilde{R}^2+\chi \tilde{R}^m$.
For $m>2$, this solution reads
\begin{eqnarray} \label{SolR2corr}
&&a=a_0 t^{\gamma}\,, \qquad \gamma=\frac{2}{3(1+w)}\,,\qquad
\mu=\frac{1-3\lambda}{3 (w-1)}\,,\nn
&& \rho=\rho_0 t^{-2}\,,\qquad \rho _0= \frac{4 (3 \lambda -1)}{3 (w+1)^2
\kappa ^2}\,.
\end{eqnarray}
This solution is obviously present only in the Ho\v{r}ava-Lifshitz version of this theory, as
one can check directly.

One of the most important methods used to investigate power law
solutions in higher order gravity is the reconstruction of a theory
starting from a specific background. In the following we will adapt
this technique to reconstruct the form of the function
$F(\tilde{R})$ that admits flat FRW power law solutions
\cite{Nojiri:2009xh,Cognola:2009za,Nojiri:2009kx,Nojiri:2006be,Nojiri:2006je,Elizalde:2010jx}.

Let us then consider a cosmological solution characterized by the Hubble
parameter
\begin{equation} \label{HubRec}
H= \frac{\gamma}{t}\,,
\end{equation}
and again assuming the energy density of a barotropic fluid
\begin{equation} \label{RhoRec}
\rho= \rho_0 t^{-3\gamma(1+w)}\, .
\end{equation}
In this case the Ricci scalar is
\begin{equation} \label{}
\tilde{R}=\frac{3 \gamma  \left(-3 \gamma  \lambda +6 \gamma  \mu +\gamma -2 \mu
\right)}{t^2}\, ,
\end{equation}
so that we can express the time $t$ as a function of $\tilde R$.  Substituting
(\ref{HubRec}) and (\ref{RhoRec}) into  (\ref{HLF13}) and (\ref{HLF16}) and
expressing $t$ in terms of $\tilde R$, one obtains
\begin{eqnarray} \label{EqRec}
&&A_1 \tilde{R}^3 F^{(3)}+A_2 \tilde{R}^2 F''+A_3 \tilde{R} F'+A_4 F+e w
\tilde{R}^{\frac{3}{2} (w+1) \gamma}=0\,,\\
&& B_1 \tilde{R}^2 F''+B_2 \tilde{R} F'+B_3 F+B_4 \tilde{R}^{\frac{3}{2}
(w+1) \gamma }=0\,,
\end{eqnarray}
with
\begin{eqnarray}
&&A_1=\frac{8 \mu }{3 \gamma  \left(\gamma-3 \gamma  \lambda +6 \gamma  \mu -2
\mu \right)},
\\&&A_2=-\frac{4 \left(\gamma -3 \gamma  \lambda +3 \mu
 \right)}{3 \gamma  \left(\gamma  \left(3 \lambda -6 \mu -1\right)+2 \mu
\right)},
\\&&A_3=\frac{2 (3 \gamma -1) \left(3 \lambda -3 \mu -1\right)}{3
\left(\gamma-3
 \gamma  \lambda +6 \gamma  \mu  -2 \mu \right)},
\\&&A_4= 3^{-\frac{3}{2} (w+1) \gamma } \kappa ^2 w \rho _0 \left[\gamma^2
(1-3  \lambda +6
   \mu ) -2 \mu\gamma\right]^{-\frac{3}{2} (w+1) \gamma }
\end{eqnarray}
and
\begin{eqnarray}
 &&B_1=\frac{4  \mu }{3 \gamma  \lambda -6 \gamma  \mu -\gamma +2 \mu },
\\&&B_2=\frac{\gamma  \left(3 \lambda -1\right)}{\gamma -3 \gamma  \lambda +6
\gamma  \mu -2 \mu   }-1,
\\&&B_3=1,
\\&&B_4=\kappa ^2 \rho _0 \left(-3^{-\frac{3}{2} (w+1) \gamma }\right)
\left(\gamma  \left(\gamma-3 \gamma  \lambda +6 \gamma  \mu  -2
\mu\right)\right)^{-\frac{3}{2} (w+1) \gamma }\,.
\end{eqnarray}
These equations admit the solution
\begin{equation} \label{SolFRec}
F(\tilde{R})=C_1 \tilde{R}^{\alpha _-}+C_2 \tilde{R}^{\alpha _+}+C_3
\tilde{R}^{\frac{3}{2} (1+w)\gamma }\, ,
\end{equation}
where
\begin{equation} \label{}
\alpha_\pm =\frac{\gamma  \left(3 \lambda -3 \mu-1\right)+3 \mu\pm\sqrt{\gamma
^2 \left(-3 \lambda +3 \mu +1\right)^2+2 \gamma  \mu  \left(3 \lambda +3 \mu
-1\right)+\mu ^2} }{4 \mu }
\end{equation}
and
\begin{equation} \label{}
C_3=\frac{3^{-\frac{3}{2} (w+1) \gamma }  \kappa ^2 \rho_0 \left[\gamma
\left(-3 \gamma  \lambda +6 \gamma  \mu +\gamma -2
\mu\right)\right]^{1-\frac{3}{2} (w+1) \gamma }}{\gamma  \left[\gamma  \left(3
\lambda -1\right) (3 (w+1) \gamma -1)-\mu  (3 (w+1) \gamma -2) (3 (w+2)
\gamma-1)\right]}\, .
\end{equation}
Note that the coefficients $\alpha$ are  real only for
$$\gamma ^2 \left(-3 \lambda +3 \mu +1\right)^2+2 \gamma  \mu  \left(3 \lambda
+3 \mu -1\right)+\mu ^2>0\, ,$$
which is satisfied for
\begin{eqnarray} \label{}
&& \gamma\geq0\\
&&\gamma<0,
\quad \mbox{and}\quad  \lambda <\frac{3 \gamma  \mu +\gamma -\mu }{3 \gamma }-2
\sqrt{-\frac{\mu ^2}{3\gamma }}\quad \lambda >\frac{3 \gamma  \mu +\gamma -\mu
}{3 \gamma } +2 \sqrt{-\frac{\mu ^2}{3\gamma}}\, .
\end{eqnarray}

\begin{figure}[htbp]
\subfigure [ The exponents of (\ref{SolFRec})  for $w=0$, $\mu=-3$ and
    $\lambda=-1/2$.]{\includegraphics[scale=1]{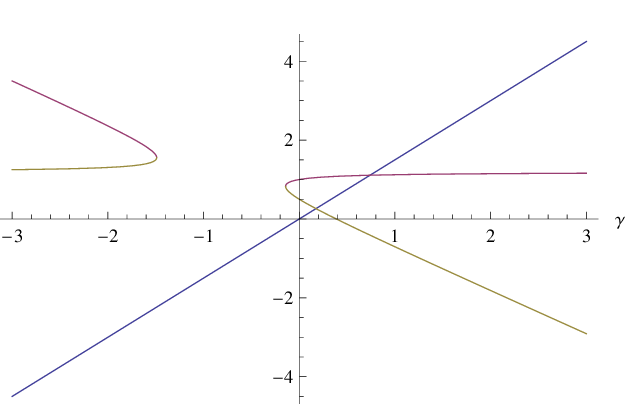}}
\subfigure[The exponents of (\ref{SolFRec})  for $w=0$, $\mu=-3$ and
$\lambda=1$.]{\includegraphics[scale=1]{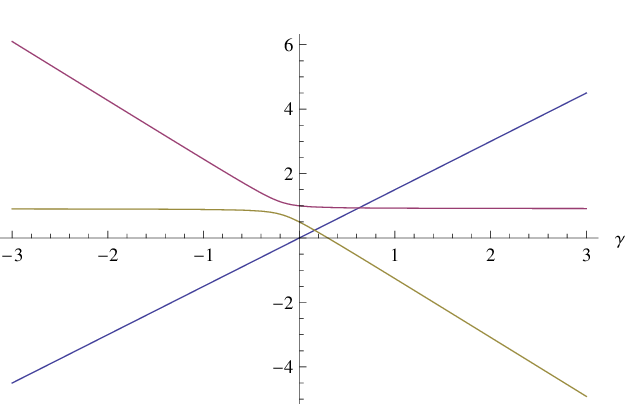}} \subfigure[ The
exponents of (\ref{SolFRec})  for $w=0$, $\mu=-2/5$ and
$\lambda=1/2$.]{\includegraphics[scale=1]{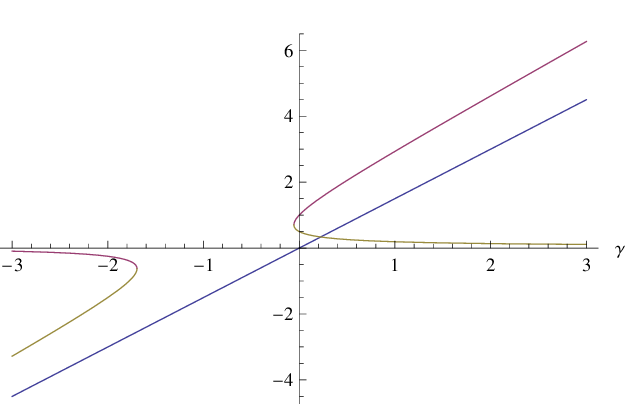}} \caption{Plot
of the curves representing the values of the exponents of the
reconstructed $F(R)$ theory corresponding to a background
$H=\gamma/t$, for $w=0$ and some specific values of the parameters 
$\lambda$ and $\mu$. From these plots one can infer, for example, 
that some backgrounds of this type can only be realized in $F(R)$ 
theories with poles.}\label{FigRec}
\end{figure}

Therefore also in the Ho\v{r}ava-Lifshitz case, the only type of function $F$
that is able to generate
analytical power law solutions is a combination of powers of the Ricci
scalar. The connection between the equations (\ref{HubRec}) and (\ref{SolFRec})
allows one to make some general considerations on the relation between the
structure of the function $F$ and the cosmology. If one plots the behavior
of the exponents of  (\ref{SolFRec}) as a function of $\gamma$ for various values of $\mu$
and $\lambda$ (see figure \ref{FigRec}),  one finds that there is a
correlation between the existence of a specific type of solutions and the
sign of the modes of the reconstructed theory. For example, if one chooses
only positive values for the exponents of (\ref{SolFRec}) in order to avoid
instabilities, none of the permitted values of the parameters are able to generate a
contracting solution. On the other hand, both a Friedmann expansion and a
power law inflation regimes can be  obtained if $\mu <0$ and $\lambda
<\frac{6 \gamma  \mu +\gamma -2 \mu }{3 \gamma
}$ or $\mu >0$ and $\lambda >\frac{6 \gamma  \mu+\gamma -2 \mu }{3 \gamma }$.
The behavior of the exponents of (\ref{SolFRec}) is shown in Figure
\ref{FigRec} for different values of the parameters.

As a final remark, it is interesting to note that the case
$F(\tilde{R})=C_2 \tilde{R}+C_3 \tilde{R}^{m}$ corresponds to the
solution (\ref{SolMat}) via the reconstruction method. This
conclusion confirms the correctness of this approach and its utility
in the search for exact solutions.

\subsection{Explicit model for the unification of inflation with dark energy}

It is interesting to try to formulate an explicit model for the 
unification of early-time inflation with late-time acceleration. One 
may propose such a model, which may unify the inflation  and the 
late-time acceleration. First we consider the case $3 \lambda -3 \mu 
-1=0$, which is specific in the Ho\v{r}ava-Lifshitz $F(R)$ gravity. 
An example is 
\be \label{F0} 
F(R) = \frac{1}{2\kappa^2} \left( 1 + c_1 \ln \kappa^2 R \right) \left(
1 - c_2 \kappa^2 R \right)\, . 
\ee 
Here $c_1$ and $c_2$ are dimensionless positive constants.
Then  following (133), 
we find two de Sitter solutions 
\be \label{F0a} R = R_L \equiv 
\kappa^2 \e^{ - \frac{1}{c_1}}\, , \quad R = R_I \equiv \frac{1}{c_2 
\kappa^2}\, . 
\ee 
If one chooses $c_1 \sim 1/280$, we find $R_L \sim 
\left( 10^{-33}\, \mathrm{eV} \right)^2$, which may describe the 
accelerating expansion of the present universe. On the other hand, 
if $c_2 \sim \mathcal{O}(1) - \mathcal{O}(100)$, $R=R_I$ may express 
the inflation.

In the general case $3 \lambda -3 \mu -1 \neq 0$, we may consider 
the following form of $F(\tilde R)$: 
\be \label{F1} 
F\left(\tilde R\right) = \tilde R + f\left(\tilde R\right)\, \quad
f\left(\tilde 
R\right) = R_I \tanh \frac{ \tilde R - R_1}{\Lambda} + R_L \tanh 
\frac{ \tilde R - R_2}{\Lambda} + R_I \tanh \frac{ R_1}{\Lambda} + 
R_L \tanh \frac{R_2}{\Lambda}\, . 
\ee 
Here $R_I$, $R_L$, $R_1$, $R_2$, and $\Lambda$ are positive constants
and we assume \be 
\label{F2} R_I \gg R_L \gg \Lambda\, \quad R_I\gg R_1\, \quad R_L 
\gg  R_2\, . \ee Then, when $\tilde R \sim R_I$, we find \be 
\label{F3} f(\tilde R) \sim R_I\, , \ee such that $f(\tilde R)$ 
plays the role of a large cosmological constant, which generates 
inflation. On the other hand, when $\tilde R \sim R_L$, $f(\tilde 
R)$ becomes a small constant, \be \label{F4} f(\tilde R) \sim R_L\, 
, \ee and the late time acceleration could be generated. Note that 
\be \label{F5} f(0) = 0\, , \ee therefore $f(\tilde R)$ is not real 
cosmological constant. Hence, explicit construction of realistic 
models for the unification of the inflation with dark energy is 
possible. The remaining freedom in the choice of parameters gives 
the possibility to make the model quite satisfactory from the 
cosmological point of view.

\subsection{Finite-time future singularities}
The attempts of constructing accelerating  cosmological models that
include a dark component have revealed that such models often
contain some unexpected phenomenology. One of the most striking
features of dark energy cosmologies is that, regardless of the way
in which  the dark component is introduced, they can become
singular. By ``become singular'' we intend that there exists a
specific time $t_s$ in which one of more key quantities of the model
becomes divergent. Some of these singularities, like the so-called
``Big Rip''  \cite{ref5}, are realized only far in the future (i.e.
$t_s>>1$). However, as it was recently pointed out 
\cite{barrow,singularity}, under special circumstances one can have
quintessence-like cosmologies that present softer singularities  at
smaller finite time (``sudden singularities''). Although in pure GR
cosmologies this pathological behavior can be cured by specifying,
for example, the equation of state of the fluid, this is not the
case in models that include dark fluids or in modified  gravity. In
fact, it has been proved  that these theories can admit up to four
different types of singularities at finite time
\cite{ConformalAnomaly1}. In the following, we will analyze the
presence of these singularities in Ho\v{r}ava-Lifshitz $F(R)$
gravity using some specific examples.

Let us define the effective density and the effective pressure associated to
the Ho\v{r}ava-Lifshitz $F(R)$ gravity. We have
\begin{eqnarray}\label{mateff}
\rho_\mathrm{eff} &=& \frac{1}{\kappa ^2}\left[6 \mu  H \dot{\tilde{R}} F''-6 \mu
\dot{H} F'+6(3\lambda  -3 \mu  -1 )H^2 F'+F\right] \,,\nn
p_\mathrm{eff} &=& \frac{1}{\kappa ^2}\left[-2 \mu  F^{(3)} \dot{\tilde{R}}^2-
2(3\lambda +\mu-1)H \dot{\tilde{R}} F''-2(3\lambda  -3 \mu  -1 ) \dot{H}
F'-6(3\lambda  -3 \mu  -1 )H^2 F'-F\right]\,.
\end{eqnarray}
The different types of finite-time singularities can be classified
by looking at the behavior of the quantities (\ref{mateff}) plus the
scale factor $a$, $H$ and its derivatives. In particular, in
\cite{ConformalAnomaly1} they are classified as
\begin{itemize}
\item Type I (``Big Rip'') : For $t \to t_s$, $a \to \infty$ and
$(\rho,|p|) \to \infty$ or $\rho$ and $|p|$ are finite;
\item Type II (``sudden'') : For $t \to t_s$, $a \to a_s$,
$\rho \to \rho_s$ and $|p| \to \infty$;
\item Type III : For $t \to t_s$, $a \to a_s$,
$\rho \to \infty$ and $|p| \to \infty$;
\item Type IV : For $t \to t_s$, $a \to a_s$
$(\rho, |p|) \to $ constant (or zero) and higher derivatives of $H$ diverge.
\end{itemize}
To classify the singularities in our case, let us consider the case of a
vacuum spatially-flat cosmology, and let us imagine that close to the time
$t_s$ the Hubble parameter can be written as
\begin{equation}\label{Hsing}
H\approx h_0(t_s-t)^{-\gamma}\,.
\end{equation}
This means that the scale factor is
\begin{equation} \label{}
a\approx a_0\exp\left[\frac{h_0(t_s-t)^{1-\gamma}}{\gamma-1}\right]\,,
\end{equation}
The above expression tells us that there are two possible behaviors
of the scale factor that depend on the value of $\gamma$: if
$\gamma\geq1$, $a$ will diverge as $t$ approaches $t_s$ while if
$\gamma<1$, $a$ will converge. Therefore it is clear that the
singularity of type I is realized when $\gamma>1$, the others for
$\gamma<1$.

The value of $\gamma$ also influences the form of the Ricci scalar. In
general, one has
\begin{equation}
\tilde{R}\approx h_0^2 \left(-9 \lambda +18 \mu +3\right)(t_s-t)^{-2 \gamma }
-6 \gamma h_0  \mu  (t_s-t)^{-(\gamma +1)}\,,
\end{equation}
but, depending on the value of $\gamma$, the above expression can be reduced
to
\begin{equation} \label{RicciSing}
\tilde{R}\approx \left\{
\begin{array}{ll}
h_0^2 \left(-9 \lambda +18 \mu +3\right) (t_s-t)^{-2 \gamma }& \mbox{if}\quad
\gamma>1\,, \\
6 h_0 \gamma  \mu  (t_s-t)^{-\gamma -1} &  \mbox{if}\quad  \gamma<1\,.  \\
\end{array}
\right.
\end{equation}
This property of the curvature also indicates that for $t\rightarrow t_s$
one has
\begin{equation}
\left\{
\begin{array}{ll}
\tilde{R}\gg 1 & \gamma>-1\,,\\
\tilde{R}\ll 1 & \gamma<-1\,,
\end{array}
\right.
\end{equation}
i.e. the curvature may become divergent or very small depending on the
value of $\gamma$. This property will turn out to be very useful for simplifying
the calculations.

Finally, because of the nature of the structure of the effective energy
density and pressure, theories with the different types of singularities are
associated directly with the values of $\gamma$. Specifically:
\begin{itemize}
\item Type I  $\Rightarrow \gamma>1$;
\item Type II $\Rightarrow -1<\gamma<0$;
\item Type III $\Rightarrow 0<\gamma<1$;
\item Type IV $\Rightarrow \gamma<-1$.
\end{itemize}
Let us now consider some simple examples in the Ho\v{r}ava-Lifshitz $F(R)$ gravity
and their comparison with the standard case.

\subsubsection{The case $F(\tilde{R})=\tilde{R}+\chi \tilde{R}^m$}
In the case
\begin{equation} \label{F=RRm}
F\left(\tilde{R}\right)=\tilde{R}+\chi \tilde{R}^m\,,
\end{equation}
substituting (\ref{Hsing}) and  (\ref{RicciSing}), one finds the
necessary conditions for the presence of the singularities:
\begin{equation} \label{singCond_beta}
\begin{array}{lc}
\mbox{Type I} &   \gamma>1
\\
\mbox{Type II} &  m<0, \quad  -1<\gamma<0\\
\mbox{Type III} & 0<\gamma<1,\quad m\neq0\\
\mbox{Type IV} &  \gamma \in \mathbb{Q}-\mathbb{Z},\quad \gamma <-1\,.
\\
\end{array}
\end{equation}
These are compatible with the solutions found in \cite{singularity}. It is
important to stress that the conditions (\ref{singCond_beta}) are only
necessary for the existence of the singularity. The reason is that we
have implicitly postulated that the solution (\ref{Hsing}) satisfies
(\ref{HLF13}) and (\ref{HLF16}) at least in a specific time interval,
which may not be the case due to the non-linearity of the theory.
Then the only way to proceed is to find some exact solutions of the
theory we are examining and see if the parameters have values
for which the conditions (\ref{singCond_beta}) are satisfied. Unfortunately
finding exact solutions in these theories can be problematic. However,
close to the singularity one can solve this problem by using the fact that, as we have seen, the magnitude of the Ricci scalar also
changes when $t \rightarrow t_s$. This means that we can approximate the
function $F$ with a simpler form that admits simple exact solutions. Then
these solutions can be used to obtain necessary and sufficient conditions
for the singularity to be realized.

In our simple example it is clear that one has
\[
F\left(\tilde{R}\right) \approx \left\{
\begin{array}{lc}
\tilde{R}^{n}    &  \gamma>-1\,,  \\
\tilde{R}  & \gamma<-1\,.
\end{array}
\right.
\]
This means that we can approximate our $F$ with $ R^{n}$ in all the cases of
interest and that
we can use the exact solutions found in Section \ref{SectionReconstruction}
in order to understand the presence of singularities.  In particular, one sees that
the solution (\ref{SolVac}) can present a singularity of Type I for
\begin{eqnarray}
&&\lambda <\frac{1}{3}\,,\qquad
\left\{
\begin{array}{cc}
0<m<\frac{1}{2}& \frac{6 m \lambda -2 m-3 \lambda +1}{8 m^3-16 m^2+16
m-8}<\mu <\frac{-6 m \lambda +2 m+3 \lambda -1}{4 m^2-12 m+8} \,,\\
\frac{1}{2}<m<1& \frac{6 m \lambda -2 m-3 \lambda +1}{6 m-6}<\mu <\frac{-6 m
\lambda +2 m+3 \lambda -1}{4 m^2-12m+8}\,,\\
1<m<2& \frac{-6 m \lambda +2 m+3 \lambda -1}{4 m^2-12 m+8}<\mu <\frac{6 m
\lambda -2 m-3 \lambda +1}{6 m-6}      \,,\\
m=2 & \mu <\frac{1}{6} (9 \lambda -3)\,,\\
m>2 & \mu <\frac{6 m \lambda -2 m-3 \lambda +1}{6 m-6}\quad \mu >\frac{-6 m
 \lambda +2 m+3 \lambda -1}{4 m^2-12 m+8} \,,
\end{array}
\right.\\
&&\lambda =\frac{1}{3}\,,\qquad m>2\qquad\mu\neq 0\\
&&\lambda >\frac{1}{3}\,,\qquad
\left\{
\begin{array}{cc}
0<m<\frac{1}{2}& \frac{-6 m \lambda +2 m+3 \lambda-1}{4 m^2-12 m+8}<\mu
<\frac{6 m \lambda -2 m-3 \lambda +1}{8 m^3-16 m^2+16 m-8}\,, \\
\frac{1}{2}<m<1& \frac{-6 m \lambda +2 m+3 \lambda-1}{4 m^2-12 m+8}<\mu
<\frac{6 m \lambda -2 m-3 \lambda +1}{6 m-6}\,,\\
1<m<2& \frac{6 m \lambda -2 m-3 \lambda +1}{6 m-6}<\mu <\frac{-6 m\lambda +2
m+3 \lambda -1}{4 m^2-12 m+8}\,,\\
m=2 & \mu >\frac{1}{6} (9 \lambda -3)\,,\\
m>2 & \mu <\frac{-6 m \lambda+2 m+3 \lambda -1}{4 m^2-12 m+8}\quad \mu>
\frac{6 m \lambda -2 m-3 \lambda +1}{6 m-6} \,,
\end{array}
\right.
\end{eqnarray}
a singularity of Type II for
\begin{eqnarray}
&&\lambda <\frac{1}{3}\,,\qquad
\left\{
\begin{array}{cc}
m<-1& 0<\mu <\frac{6 m \lambda -2 m-3 \lambda +1}{4 m^2-4}\,,\\
m=-1& \mu >0\,,\\
-1<m<0& \mu <\frac{6 m \lambda -2 m-3 \lambda +1}{4 m^2-4}\qquad \mu >0\,,
\end{array}
\right.\\
&&\lambda =\frac{1}{3}\,,\qquad -1<m<0\qquad \mu\neq 0\\
&&\lambda >\frac{1}{3}\,,\qquad
\left\{
\begin{array}{cc}
m<-1 &  \frac{6 m \lambda -2 m-3 \lambda +1}{4 m^2-4}<\mu<0\,,\\
m=-1& \mu <0\,, \\
-1<m<0& \mu <0 \qquad \mu >\frac{6 m \lambda -2 m-3 \lambda +1}{4 m^2-4} \,,
\end{array}
\right.
\end{eqnarray}
and a singularity of Type III for
\begin{eqnarray}
&&\lambda <\frac{1}{3}\,,\qquad
\left\{
\begin{array}{cc}
m<0& \mu <\frac{6 m \lambda -2 m-3 \lambda +1}{6 m-6}\,,\\
0<m<\frac{1}{2}& \frac{-6 m \lambda +2 m+3 \lambda -1}{4 m^2-12 m+8}<\mu
<0\,,\\
\frac{1}{2}<m<1& \mu <0\quad \mu >\frac{-6 m \lambda +2 m+3 \lambda -1}{4
m^2-12 m+8}\,,\\
1<m<2& \mu<\frac{-6 m \lambda +2 m+3 \lambda -1}{4 m^2-12 m+8}\quad \mu
0\,,\\
m=2 &  \mu >0\,, \\
m>2 & 0<\mu <\frac{-6 m \lambda +2 m+3 \lambda -1}{4 m^2-12 m+8}\,,
\end{array}
\right.\\
&&\lambda =\frac{1}{3}\,,\qquad \frac{1}{2}<m<2\qquad m\neq1\qquad\mu\neq
0\\
&&\lambda >\frac{1}{3}\,,\qquad
\left\{
\begin{array}{cc}
m<0 & \mu <0\quad \mu >\frac{6 m \lambda -2 m-3 \lambda +1}{6 m-6}\,,\\
0<m<\frac{1}{2}& 0<\mu <\frac{-6 m \lambda +2 m+3\lambda -1}{4 m^2-12
m+8}\,, \\
\frac{1}{2}<m<1& \mu <\frac{-6 m \lambda +2 m+3 \lambda -1}{4 m^2-12
m+8}\qquad \mu>0\,,\\
1<m<2& \mu <0\quad \mu >\frac{-6 m \lambda +2 m+3 \lambda -1}{4 m^2-12
m+8}\,,\\
m=2 & \mu <0\\
m>2 & \frac{-6 m \lambda +2 m+3 \lambda -1}{4 m^2-12 m+8}<\mu <0 \,.
\end{array}
\right.
\end{eqnarray}
This is very different from the standard case, where we have a singularity
of Type I only for $m>2$, and a Type III for $\frac{1}{2}<m<1$ and $m<0$.

For the solution  (\ref{SolMat})  we have a singularity of Type III when
$$\left\{
\begin{array}{ccc}
-3<m\leq -\frac{3}{2}& \frac{1}{3} (-2 m-3)<w\leq 1\,,\\
-\frac{3}{2}<m<\frac{1}{2}\left(1-\sqrt{13}\right)& 0\leq w\leq 1\,.
\end{array}
\right.$$

For the new solution (\ref{Sol HL}), which exists only in the Ho\v{r}ava-Lifshitz version of
$F(R)$ gravity, we have instead a singularity of Type I for
$$\left\{
\begin{array}{ccc}
\lambda <\frac{1}{3}&\frac{1}{6} (3 \lambda -1)<\mu <\frac{1}{8} (3 \lambda
-1)&m>1\,,\\
\lambda >\frac{1}{3}&\frac{1}{8} (3 \lambda -1)<\mu <\frac{1}{6} (3 \lambda
-1)& m>1\,,
\end{array}
\right.$$
and a Type III for
$$\left\{
\begin{array}{ccc}
\lambda <\frac{1}{3}& \frac{1}{8} (3 \lambda -1)<\mu <0 & m>1\,,\\
\lambda >\frac{1}{3} & 0<\mu <\frac{1}{8} (3 \lambda -1)& m>1\,.
\end{array}
\right.$$
The results above show clearly that the presence of singularities is deeply
altered in the Ho\v{r}ava-Lifshitz version of $F(R)$ gravity.  In particular, it seems
that the additional parameters make it much easier to realize the singularities.
The intervals we have presented above for the parameters can then be
interpreted as constraints on this type of Ho\v{r}ava-Lifshitz $F(R)$ theories of gravity. Thus,
we have demonstrated that modified Ho\v{r}ava-Lifshitz gravity has the
phantom-like or quitessence-like accelerating cosmologies, which might
lead to singularities of type I, II, or III.

\subsubsection{Eliminating the singularities}

Using conformal techniques, it was first argued in \cite{Abdalla:2004sw}
that in the standard $F(R)$ gravity the singularities of the type we have found
can be cured, when additional powers of the Ricci scalar are added to the
Lagrangian. We can then verify if  something similar happens also in the
Ho\v{r}ava-Lifshitz case. Let us therefore consider the case
\begin{equation} \label{F=RR2Rm}
F(\tilde{R})=\tilde{R}+\xi \tilde{R}^2+\chi \tilde{R}^m\,.
\end{equation}
The action of this theory contains a correction of order $\tilde{R}^2$ and,
in the standard case, it is able to cure the singularities of  theories of
the type $\tilde{R}+\chi \tilde{R}^m$.

If one derives the necessary conditions for the presence of a singularity,
one finds:
\begin{equation} \label{}
\begin{array}{lc}
\mbox{Type I} &   \gamma>1,
\\
\mbox{Type II} &  \mbox{never}, \\
\mbox{Type III} & 0<\gamma<1\quad m\neq0\,,\\
\mbox{Type IV} &  \gamma \in \mathbb{Q}-\mathbb{Z}\quad \gamma <-1 \,.
\\
\end{array}
\end{equation}
One can already see at this level that in this kind of theory singularities
of Type II never occur. This can be interpreted as the fact  that the
correction $R^2$ is able to compensate for these terms.

As said before, the conditions above are only necessary and the only way
to actually determine if a singularity is present is to consider
an exact solution of the theory and see if the conditions above can be
satisfied by that solution. Let us consider then the solution
(\ref{SolR2corr}) found in the previous section. Applying the conditions
above one obtains that none of them is satisfied for this background. In
other words, the addition of the $\tilde{R}^2$ term compensates for the
singularities one would find in a similar background of (\ref{F=RRm}). This
supports the claim made in  \cite{Abdalla:2004sw} that (\ref{F=RR2Rm}) is
more regular than (\ref{F=RRm}) and  that in general the introduction of
additional curvature invariants into the action can help in the curing of the
singularities of an $F(R)$ theory of gravity. On the other hand, it is also
quite possible that the account of quantum gravity effects
\cite{Elizalde:2004mq} may also cure the future singularities.

Summarizing, because of the additional parameters, the Ho\v{r}ava-Lifshitz
version of $F(R)$ gravity has a bigger space of de Sitter solutions
compared to its standard counterpart.
Using a simple theory and both a direct resolution of the cosmological
equations and a reconstruction technique,  we have also
verified that  this is true in the case of power law solutions. In general
the presence of additional parameters also means that one has a bigger freedom
in the choice of the features of these solutions, e.g. to see if it is possible
to realize accelerated expansion.  Therefore in these theories the realization
of dark energy era (and inflation) is comparatively easier.
This is interesting because it draws a direct connection between the Newtonian
nature of quantum gravity and the observed behavior of the Universe.
Unfortunately, the additional number of parameters also increases
the probability that these cosmologies will  become singular not only at
$t\to \infty$, but also at finite time. Using exact solutions, we have shown that
there are many combinations of values of the parameters of the theory which
are able to induce the appearance of singularities. We have also verified,
via a specific example, that adding an invariant of the type $R^2$ into
the Lagrangian of a theory, one is able to obtain a theory whose solutions
are much more regular. Consequently, also in the Ho\v{r}ava-Lifshitz
case one can compensate such ill behavior of these models by the
introduction of additional powers of the Ricci scalar into the action.

\section{Corrections to the Newton law}
\label{sec:7}

Let us now consider the possible corrections to the Newton law.
For this purpose, we consider the infrared region where the higher
derivative terms like
(\ref{HLFrg9}) or (\ref{HLFrg14}) can be neglected and we find
\be
\label{KLFrg17}
\mathcal{L}_R^{(3)} \left(g^{(3)}_{ij}\right)
\sim R^{(3)}\, .
\ee
The $R^{(3)}$ term can be added from the beginning or this term might be
induced at the infrared fixed
point \cite{Horava:2009uw}.
Then by the transformation (\ref{HLFrg4}), by using (\ref{HLFrg10})
and (\ref{HLFrg5}),
in (\ref{HLFrg1b}), one gets
\bea
\label{KLFrg18}
&& \int \rd^4 x \sqrt{{\bar g}^{(3)}} \left\{
\bar{\mathcal{L}}_R^{(3)}\left( {\bar g}^{(3)}_{ij}, \varphi \right)
 - V(\varphi) \right\} \nn
&& = \int \rd^4 x \sqrt{{\bar g}^{(3)}} \left\{
\e^\varphi \left( {\bar R}^{(3)}
 - \frac{5}{2} {\bar g}^{(3) kl} {\bar \nabla}^{(3)}_k \varphi
{\bar \nabla}^{(3)}_l \varphi \right)
 - \left(A\left(\varphi\right)F' \left(A\left(\varphi\right)\right))
 - F\left(A\left(\varphi\right)\right)\right) \right\} \, .
\eea
The usual Newton law can be generated through the exchange of the graviton.
Furthermore by the exchange of
the scalar field $\varphi$, extra force might be generated.
Now we consider the case that
\be
\label{KLFrg19}
F(A) = A - \Lambda_\mathrm{eff} - \frac{c}{A^n} +
\mathcal{O}\left(A^{-n-1}\right)\, .
\ee
Here $\Lambda_\mathrm{eff}$ is an effective cosmological constant.
Then
\be
\label{KLFrg20}
F'(A) = 1 + \frac{cn}{A^{n+1}} + \cdots \, .
\ee
In the solar system or on the earth, the second term in
(\ref{KLFrg20}) is much smaller than unity,
which corresponds to the first term.
Hence, we find
\be
\label{KLFrg21}
\varphi\equiv \frac{1}{3}\ln F'(A) \sim \frac{cn}{3A^{n+1}} \, ,
\ee
and therefore
\be
\label{KLFrg22}
V(\varphi) \sim
\Lambda_\mathrm{eff} + \frac{c(n+1)}{A^n} \sim \Lambda_\mathrm{eff}
+ \frac{\left(n+1\right) c^{\frac{1}{n+1}}}{n^{\frac{n}{n+1}}}
\varphi^{- \frac{n}{n+1}} \, .
\ee
Then since $\e^\varphi\sim 1$, the effective mass $m_\varphi$ of
$\varphi$ is given by
\be
\label{KLFRg23}
m_\varphi^2 \equiv \frac{V''(\varphi)}{5} \sim
\frac{n^{\frac{1}{n+1}}(2n+1) c^{\frac{1}{n+1}}}{n+1}
\varphi^{- \frac{3n+2}{n+1}}
\sim \frac{3^{\frac{3n+2}{n+1}} (2n+1)}{n^{\frac{3n+1)}{n+1}}(n+1)
c^{\frac{3n+1}{n+1}}} A^{3n+2}\, .
\ee
In the ``realistic'' model, $c$ is chosen to be $c=\mu^{2(n+1)}$ and
$\mu\sim 10^{-33}\,\mathrm{eV}$.
On the other hand, we find $A=R \sim 10^{-61}\,\mathrm{eV}^2$ in the
solar system and
$A=R \sim 10^{-50}\,\mathrm{eV}^2$ on the earth. Thus,
one finds $m_\varphi^2 \sim 10^{15n - 56}\,\mathrm{eV}^2$ in the
solar system and
$m_\varphi^2 \sim 10^{48n - 34}\,\mathrm{eV}^2$. Hence, if
$n$ could be large enough, the mass
of $\varphi$ would become large and the Compton length would become
small, so that the correction to the Newton law would not be observed.

\section{Discussion and conclusions}
\label{sec:8}

We have proposed a first-order modified Ho\v{r}ava-Lifshitz-like
gravity action and studied its Hamiltonian structure. As a large
explicit class of such models we considered the modified
Ho\v{r}ava-Lifshitz $F(R)$ gravity that is more general than the one
introduced in ref.~\cite{Chaichian:2010yi}, which for the special
choice of parameter $\mu=0$ coincides with the degenerate model
introduced in ref.~\cite{Kluson:2009xx}. Its ultraviolet properties
are discussed and it is demonstrated that such $F(R)$ gravity may be
renormalizable for the case $z=3$ in a similar way as the original
proposal for Ho\v{r}ava-Lifshitz gravity. The Hamiltonian analysis
of the proposed modified Ho\v{r}ava-Lifshitz $F(R)$ gravity shows
that this theory is generally consistent with reasonable
assumptions. The $F(R)$ gravity action has also been analyzed in the
fixed gauge form, where the presence of the extra scalar is
particularly illustrative. The methods presented in the Hamiltonian
analyses of sections \ref{sec:3} and \ref{sec:4} can be used to
study any action of the general form (\ref{HLF26}).

The spatially-flat FRW cosmology of the modified Ho\v{r}ava-Lifshitz
$F(R)$ gravity is studied. It is shown that it coincides with the
one of the earlier model \cite{Chaichian:2010yi}, but only in the
spatially-flat FRW case. For specific choice of the parameters of
the theory, its FRW equations of motion coincide with the ones of
the traditional $F(R)$ gravity. The presence of the multiple de
Sitter solutions shows the principal possibility of the unification
of the early-time inflation with the late-time acceleration in the
Ho\v{r}ava-Lifshitz background, which proves that it can have rich
cosmological applications. The power-law theories are investigated
in detail. A number of analytical FRW solutions is found, including
the ones with behavior relevant for the early/late cosmic
acceleration. The quintessence/phantom-like cosmologies derived in
our work may show all the four possible types of finite-time future
singularities like in the case of standard dark energy. The
conditions to cure such future singularities are discussed in
analogy with the traditional $F(R)$ gravity. It is also interesting
that the correction to the Newton law in the $F(R)$ gravity under
discussion can be made unobservably small. Finally, a covariant
proposal for $F(R)$ gravity in Ho\v{r}ava-Lifshitz spirit has been
made.

Despite some successes in the formulation of modified
Ho\v{r}ava-Lifshitz $F(R)$ gravity which can be made renormalizable
and in its cosmological applications, a number of unsolved questions
remain. What is the appropriate way to introduce matter in the
theory? Is the theory itself fundamental (or at least, fully
consistent) or does it descend from another more fundamental
proposal? Can it comply with all the local tests in the Solar system
as well as with cosmological bounds? What is the dynamical scenario
for the restoration of the Lorentz invariance at late times? What
are the cosmological and astrophysical consequences of the
first-order modified Ho\v{r}ava-Lifshitz gravity when compared with
those of the traditional modified gravity \cite{tradModGrav}.
Moreover, the traditional questions about the properties of black
holes in such a theory can be straightforwardly investigated.

Nevertheless, even at the present stage some surprises can be expected from
the theory.

While the universe has likely undergone a perioid of inflation in its 
early moments, it is interesting to note that Ho\v{r}ava-Lifshitz 
gravity could produce cosmological perturbations that are  almost 
scale-invariant even without inflation \cite{Mukohyama:2009gg}.

Ho\v{r}ava-Lifshitz gravity has also been considered in the presence 
of scalar fields \cite{Chen:2009ka,Lee:2010iu}.
In principle, it is possible to extend our Ho\v{r}ava-Lifshitz $F(R)$ 
gravity by including its coupling with scalar fields.

We would also like to mention a recent paper \cite{Kluson:2010aw},
where a new class of Lorentz-invariance breaking non-relativistic
string theories, inspired by the Ho\v{r}ava-Lifshitz gravity, has
been presented and analyzed.

Using the $F(R)$ version of gravity one can propose even a more
general formulation of string theory in the Ho\v{r}ava-Lifshitz
background: for instance, rigid strings, membranes and $p$-branes,
etc. On the other hand, it may suggest unusual solutions for the
known cosmological problems. There also exists an attempt to explain
the homogeneity of our universe in a model with varying speed of
light \cite{Albrecht:1998ir}. Having in mind that in the ultraviolet
region the speed of the Ho\v{r}ava-Lifshitz graviton changes, one
may speculate that the homogeneity of the universe may be described
without the need for inflation. In any case, such a theory is both
theoretically and cosmologically rich and it deserves further study.

\section*{Acknowledgments}

This research has been supported in part by MEC (Spain) project
FIS2006-02842 and AGAUR (Catalonia) 2009SGR-994 (SDO), by Global COE
Program of Nagoya University (G07) provided by the Ministry of
Education, Culture, Sports, Science \& Technology (SN). M. O. is
supported by the Finnish Cultural Foundation. The support of the
Academy of Finland under the Projects No. 121720 and 127626 is
gratefully acknowledged.

\appendix
\section{Proposal for a covariant $F(R)$ gravity}
\label{appendix}

In \cite{Nojiri:2009th}, a new type of covariant
Ho\v{r}ava-Lifshitz-like gravity has been proposed.
The action is given, for $z=2$ case
\be
\label{Hrv1}
S = \int d^4 x \sqrt{-g} \left\{ \frac{R}{2\kappa^2} - \alpha \left(
T^{\mu\nu} R_{\mu\nu}
+ \beta T R \right)^2 \right\}\, ,
\ee
and for $z=3$ case
\be
\label{Hrv15}
S = \int d^4 x \sqrt{-g} \left\{ \frac{R}{2\kappa^2} - \alpha \left(
T^{\mu\nu} R_{\mu\nu}
+ \beta T R \right)
\left(T^{\mu\nu}\nabla_\mu \nabla_\nu + \gamma T \nabla^\rho
\nabla_\rho\right)
\left( T^{\mu\nu} R_{\mu\nu} + \beta T R \right)
\right\}\, ,
\ee
and for $z = 2 n + 2$ case,
\be
\label{Hrv18}
S = \int d^4 x \sqrt{-g} \left[ \frac{R}{2\kappa^2} - \alpha \left\{
\left(T^{\mu\nu}\nabla_\mu \nabla_\nu + \gamma T \nabla^\rho
\nabla_\rho\right)^n
\left( T^{\mu\nu} R_{\mu\nu} + \beta T R \right) \right\}^2 \right]\
.
\ee
Here $T_{\mu\nu}$ is the energy-momentum tensor of the perfect fluid
with constant equation
of state parameter $w$ and the parameters $\beta$ and $\gamma$ are
given by
\be
\label{Hrv16}
\beta = - \frac{w-1}{2\left(3w - 1\right)}\, ,\quad
\gamma = \frac{1}{3 w - 1}\ .
\ee
Note that this fluid is not the standard matter fluid. It may have
a stringy origin or it may be a kind of gravitational fluid.
Hence, we may consider the following covariant
Ho\v{r}ava-Lifshitz-like gravity:
\bea
\label{KLFRg24}
S_{F(\tilde R_\mathrm{cov} )} &=& \frac{1}{2\kappa^2}\int \rd^4 x
\sqrt{- g} F(R_\mathrm{cov})\, , \nn
R_\mathrm{cov} &=& \left\{
\begin{array}{ll}
R - 2 \alpha \kappa^2 \left( T^{\mu\nu} R_{\mu\nu}
+ \beta T R \right)^2\, , & z=2 \, ,\\
R - 2\kappa^2 \alpha \left( T^{\mu\nu} R_{\mu\nu}
+ \beta T R \right)
\left(T^{\mu\nu}\nabla_\mu \nabla_\nu + \gamma T \nabla^\rho
\nabla_\rho\right)
\left( T^{\mu\nu} R_{\mu\nu} + \beta T R \right) & z=3 \, ,\\
R - 2\kappa^2 \alpha \left( T^{\mu\nu} R_{\mu\nu}
+ \beta T R \right)
\left(T^{\mu\nu}\nabla_\mu \nabla_\nu + \gamma T \nabla^\rho
\nabla_\rho\right)
\left( T^{\mu\nu} R_{\mu\nu} + \beta T R \right) & z = 2 n + 2 \, .\\
\end{array}
\right. \eea The theories given by the action (\ref{KLFRg24}) could
be renormalizable when $z\geq 3$. This may be demonstrated using the
same arguments as in ref.~\cite{Nojiri:2009th}. This is a large
class of covariant $F(R)$ gravities whose ultraviolet properties are
more similar to the ones of Ho\v{r}ava-Lifshitz-like gravity.
However, in many respects their cosmology is similar to the
cosmology of traditional modified theories of gravity
\cite{Nojiri:2006ri}.

\end{document}